\documentclass[a4paper,11pt]{JHEP3}

\usepackage{preamble}

\begin{document}

\section{Introduction}

We model spin-1 resonances and the massless pions of QCD in the chiral limit.
The aim is to improve the understanding of the interplay between low-energy
hadron physics and the known behavior of perturbative QCD. For this purpose,
we build and explore a model of $\tmop{SU} \left( N_f \right) \times \tmop{SU}
\left( N_f \right)$ Yang-Mills (YM) fields with a compact extra dimension. We stress that the fifth dimension is used as a tool here, and is not meant to be physical. From the four-dimensional (4D) point of view, the
spectrum contains an infinite number of resonances, as expected in large-$N_c$
QCD {\cite{'tHooft:1974jz,Witten:1979kh}}. Note that we limit ourselves to $\tmop{SU} \left( N_f \right)$ rather
than $\mathrm{U} \left( N_f \right)$ for simplicity.

To discuss a model of mesons, it is appropriate to first take into account the
symmetry constraints on the lightest mesons: the pions, which are the
Goldstone bosons (GBs) of the spontaneously broken 4D
$\tmop{SU} \left( N_f \right) \times \tmop{SU} \left( N_f \right)$ chiral
symmetry. These symmetry constraints are embodied by Chiral Perturbation
Theory ($\chi$PT) {\cite{Weinberg:1979kz,Gasser:1984yg,Gasser:1985gg}}. To go
beyond the range of validity of this effective theory, i.e. to energies higher
than the mass of the $\rho$ resonance, one resorts to models of resonances:
various models exist for the $\rho$ and $a_1$
{\cite{Bando:1984ej,Meissner:1988ge,Bando:1988ym,Bijnens:1992uz}}. Besides
pions, these resonances play the leading role in low-energy hadron physics
{\cite{Ecker:1989te}}. Their couplings are constrained by requiring matching
with the high-energy theory {\cite{Ecker:1989yg}}.

In order to recover perturbative QCD logarithms in two-point functions, one
needs a model with either an infinite number of resonances or a
perturbative continuum. Many efforts have been devoted recently to explore the
constraints imposed on these models
{\cite{Peris:1998nj,Shifman:2000jv,Golterman:2001nk,Golterman:2001pj,Golterman:2002mi,Afonin:2004yb,Cata:2005zj,Sanz-Cillero:2005ef}}
by the matching with both $\chi$PT at low energies and QCD at high energies.
Namely, this tells us about the symmetries, the soft high-energy behavior and
the bearing of QCD at large-$N_c$ on the model.

A five-dimensional (5D) model naturally describes such an infinite number of
states: with a compact extra dimension, each 5D field can alternatively be
seen as a tower of Kaluza-Klein (KK) excitations. We are interested in such an
infinite tower of massive spin-1 fields, supposedly describing the vector and
axial resonances ($\rho, a_1, \rho', \cdots$)
{\cite{Son:2003et,Sakai:2004cn}}. Having a lagrangian formalism allows us to
go beyond two-point functions. In addition, the a priori infinite number of
resonance parameters are fixed by only one gauge coupling and the geometry,
described by one additional parameter. These two parameters correspond to
$N_c$ and $\Lambda_{\tmop{QCD}}$.

The fact that the spin-1 mesons come from the same 5D YM field has deeper
implications than just labeling the resonances. In the 5D model, higher-dimensional gauge invariance is realized among the meson interactions not only
as infinitely many (higgsed) 4D gauge invariances, but it also contains a
residual symmetry {\cite{Arkani-Hamed:2001nc}}, whose GBs describe the pions.
Since the pions descend from the original YM field, their interactions with
mesons are constrained. As we show in this paper, this ensures certain
relations constraining the high-energy behavior of various amplitudes.

To understand the origin of this soft high-energy behavior, it is useful to go back to
those 4D models which introduced the $\rho$ meson as a YM field
{\cite{Bando:1984ej,Meissner:1988ge}}. As shown in {\cite{Ecker:1989yg}}, such
models automatically exhibit a soft high-energy behavior for some amplitudes, as
required by QCD. The models could be generalized to include the $a_1$ meson
{\cite{Bando:1988ym}}. Using gauge fields at the sites of a
latticized/deconstructed extra-dimension {\cite{Arkani-Hamed:2001ca}}, one can
even describe an infinite tower of vector and axial mesons
{\cite{Son:2003et}}. However, there is a fundamental ingredient added in the
5D gauge theory: exact {\tmem{locality}} in the extra dimension. This implies
an even softer high-energy behavior {\cite{{Barbieri:2003pr,Hirn:2004}}.

An independent motivation to consider a 5D model with YM fields comes from the
AdS/CFT correspondence
{\cite{Maldacena:1997re,Gubser:1998bc,Witten:1998qj}} (see also
refs.
{\cite{Arkani-Hamed:2000ds,Rattazzi:2000hs,Perez-Victoria:2001pa}} for
an early field-theory approach in the Randall-Sundrum model).
This correspondence established an equivalence between two very precise
theories, one defined in 4D, and the other in 5D. None of them has any known
phenomenological applications. On the other hand, many efforts have been done to find a dual description of QCD  (see among others {\cite{Karch:2002sh,Karch:2002xe,Kruczenski:2003be,Hong:2003jm,Evans:2004ia,Paredes:2004is}}). The bet here is to take another approach, using a simple idea behind 
AdS/CFT: {\tmem{a global symmetry in 4D corresponds to its gauged version
realized in the 5D bulk}}, and to see how a 5D YM model can mimic QCD.

Notice that the soft high-energy behavior given by the 5D YM model may be welcome from
the point of view of QCD. This is the case for the vector form factor,
as we show in this paper. When undesirable, the too-soft high-energy behavior should be corrected in order to agree with QCD. For instance, 5D locality implies an
infinite set of vanishing sum rules {\cite{Barbieri:2003pr,Hirn:2004ze}},
generalizing those known as Weinberg sum rules (WSRs) in QCD
{\cite{Weinberg:1967kj,Shifman:1979bx,Knecht:1998ts}}. This high-energy behavior is
even softer than in QCD, in which the presence of a quark-antiquark condensate
leaves us with only two vanishing Weinberg sum rules (in the chiral limit).
This discordance can be cured by the adjunction of scalar degrees of freedom in the bulk
(corresponding to the quark condensate and its spin-0 excitations)
{\cite{Hirn:2004,Erlich:2005qh,daRold:2005zs}}. We avoid this complication
here: while still retaining many features of QCD and a good agreement with
phenomenology, this allows us to push analytical computations further. We also
propose a way of introducing the $\left\langle \overline{q} q \right\rangle$
condensate without additional degrees of freedom.

The simplicity of our model comes from the way the spontaneous breaking of the
4D global $\tmop{SU} \left( N_f \right) \times \tmop{SU} \left( N_f \right)$
symmetry is implemented: by identifying gauge connections away from our world,
i.e. at a distance in the fifth dimension. This breaking can be communicated
to our world only by infrared effects, and will only generate
{\tmem{non-local}} order parameters of chiral symmetry. The virtue is that the
5D YM fields yield a multiplet of pions coming from the fifth component of the
axial vector field $A_5$, without the need for any field other than 5D YM.

This is different from the 5D model of {\cite{Erlich:2005qh,daRold:2005zs}},
which takes into account a quark condensate, and in which bulk pseudo-scalars
mix with the $A_5$. The inclusion of quark condensate, quark masses and spin-0
resonances in our model is deferred to future work. Even without these
ingredients, we have the usual set of {\tmem{non-local}} order parameters,
with non-vanishing values given by the low energy constants of the chiral
lagrangian: the $L_i$'s, and first and foremost, $f_{\pi}^2$. In our model,
{\tmem{all}} local order parameters, including $\left\langle \overline{q} q
\right\rangle$, are treated on an equal footing: they must be introduced
separately in the model as infrared (IR) modifications of the metric
{\footnote{Note that, as in the model of {\cite{Erlich:2005qh,daRold:2005zs}},
an {\tmem{independent parameter}} has to be introduced for the quark
condensate. This has no equivalent in QCD.}}. We present a general method to
introduce condensates, whether order parameters or not.

From the 4D point of view, our treatment is also different. After describing
the setup in Section \ref{setup}, we introduce in Section \ref{field-redef} a
specific implementation of the holographic recipe that separates explicitly
between sources and fields. In Section \ref{GBs}, we pay particular attention
to the chiral symmetries and to the matching with very low energies; i.e.
$\chi$PT.

Regarding intermediate energies (Section \ref{resonances}), we obtain the
resonance spectra and couplings. These parameters satisfy sum rules: in
Section \ref{HE} we show that these result in a soft high-energy behavior, as expected in QCD.
Two-point functions of quark currents are considered in Section \ref{2point}:
the partonic logarithm is recovered. We show how to
reproduce the condensates in the OPE without introducing additional degrees of
freedom. We also explain in what sense the model gives a prescription for
handling infinite sums over resonances in Appendix \ref{AD}.

\section{The setup} \label{setup}

The model will contain the essential features in order to interpolate between
the known behavior at low and high energies of Green's functions of the QCD
currents. It is directly formulated in terms of mesons, and never refers to
quarks. Our {\tmem{inputs}} are the following: (i) a five-dimensional
space-time and (ii) 5D $\tmop{SU} ( N_f ) \times \tmop{SU} ( N_f )$ gauge
symmetry, broken by boundary conditions. 

The {\tmem{outputs}} will be: (i) spontaneously broken 4D $\tmop{SU} \left(
N_f \right) \times \tmop{SU} \left( N_f \right)$ symmetry and its GBs, the
pions, (ii) an infinite number of vector and axial resonances with a clear
prescription for handling them and, (iii) soft behavior at high energies.

If one further asks for conformal invariance in part of the geometry, partonic logarithms can be recovered.

We write the 5D metric as
\begin{eqnarray}
  \mathd s^2 & = & g_{M N} \mathd x^M \mathd x^N, \hspace{2em} \text{with } M,
  N = 0, 1, 2, 3, 5, 
\end{eqnarray}
where we will denote the coordinates as $( x^{\mu}, z )$ with $\mu = 0,
\cdots, 3$ and $x^5 \equiv z$ representing the position in the fifth
dimension. The signature is $\left( + - - - - \right)$. We will keep general
covariance in most of the computations, but always assume that the metric can
be put in the {\tmem{conformally flat}} form
\begin{eqnarray}
  \mathd s^2 & = & w \left( z \right)^2  \left( \eta_{\mu \nu} \mathd x^{\mu}
  \mathd x^{\nu} - \mathd z^2 \right) .  \label{2.2}
\end{eqnarray}
Such a metric enjoys 4D Poincaré invariance for each hypersurface of constant
$z$. The extra dimension considered here is an interval {\footnote{The
interval is more suitable than the orbifold for our purposes \cite{vonGersdorff:2004cg}. Indeed, in
Section \ref{field-redef}, we will need to consider more general boundary
conditions than the ones used in the orbifold approach: the values of the
fields at one boundary will play the role of classical sources for the
generating functional.}}. The two ends of the space are located at $l_0$ (the
{\tmem{UV brane}}) and $l_1$ (the {\tmem{IR brane}}), with the names UV/IR
implying $w \left( l_0 \right) \geqslant w \left( l_1 \right)$  for the warp factor of (\ref{2.2}).

\subsection{The symmetries and the symmetry-breaking pattern}

We wish to study Green's functions of the quark currents of QCD. For this, we
construct a model of mesons. The lagrangian, as well as the expressions for
the currents will involve only bosons. The symmetry-breaking pattern
can be understood as  a {\it folded} version of the one in
\cite{Son:2003et}. As will be seen at the end of the paper, this
folding is essential for the introduction of local order parameters,
absent in \cite{Son:2003et}.

\subsubsection{5D $\tmop{SU} ( N_f ) \times \tmop{SU} ( N_f )$ gauge
invariance}

The global 4D chiral symmetry is upgraded to a gauge symmetry in the bulk, whose
particular realization at one special point $z = l_0$ (UV brane) has to be
identified with the chiral symmetry of QCD. As already mentioned in the
Introduction, this 5D gauged version of the chiral symmetry finds its origin
in the AdS/CFT correspondence. This fixes the field content: we consider a 5D
$\tmop{SU} \left( N_f \right) \times \tmop{SU} \left( N_f \right)$ gauge
theory.

The model includes in one step an infinite number of resonances: they are the
4D excitations of the same 5D YM field. Regarding the lagrangian, we limit
ourselves in this paper to operators involving no more than two derivatives,
and work at tree level {\footnote{The hope is that a consistent treatment at
the level of loops can be formulated as an effective theory. The action would
then have to include operators with more derivatives: derivative and loop
expansion should be related, using power-counting rules such as those in the
Appendices of {\cite{Hirn:2005fr}}.}}. Therefore, the description is
completely fixed once one specifies the 5D gauge coupling and the metric: the
action is
\begin{eqnarray}
  \mathcal{S}_{\tmop{YM}} & = & - \frac{1}{4 g_5^2}  \int \mathd^4 x \int_{l_0}^{l_1}
  \mathd z \sqrt{g} g^{M N} g^{R S}  \left\langle L_{M R} L_{N S} + R_{M R}
  R_{N S} \right\rangle,  \label{action}
\end{eqnarray}
where $\left\langle \cdots \right\rangle$ means the trace in flavor space, and
$R_{M N} \equiv \partial_M R_N - \partial_N R_M - \mathi [ R_M, R_N ]$. The
square of the 5D YM coupling $g_5^2$ has dimensions of length. The action
(\ref{action}) is invariant under ``parity'' $L \leftrightarrow R$ and the 5D
$\tmop{SU} \left( N_f \right) \times \tmop{SU} \left( N_f \right)$ gauge
transformations denoted by $R \left( x, z \right), L \left( x, z \right)$
acting as $R_M \equiv R_M^a \frac{T^a}{\sqrt{2}} \longmapsto RR_M R^{\dag} + \mathi R
\partial_M R^{\dag}$, with the generators of $\tmop{SU}\left(N_f\right)$ normalized by $\left\langle T^a T^b\right\rangle= 2 \delta^{a b}$.

The 5D gauge fields $L_M $ and $R_M$ enjoy a higher-dimensional gauge
invariance. From the 4D point of view, 5D gauge invariance is seen as an
infinite number of 4D gauge invariances plus a {\tmem{shift symmetry}}
{\cite{Arkani-Hamed:2001nc,vonGersdorff:2002rg}}. In Section \ref{thea5} we
will make use of this residual 4D symmetry to identify the pions. In QCD, the
pions are the GBs of chiral symmetry breaking. Here, the pions will be the GBs
of the shift symmetry, the remnant of the 5D gauge symmetry.

\subsubsection{Symmetry breaking by BCs}

We break the symmetry by boundary conditions, which distinguish between the
vector and axial combinations of the $L_M$ and $R_M$ fields. This will produce
the spontaneous breaking of chiral symmetry at low energies. We impose at the
IR brane
\begin{eqnarray}
  R_{\mu} \left( x, z = l_1 \right) - L_{\mu} \left( x, z = l_1 \right) & = &
  0 .  \label{R-LIR}
\end{eqnarray}
This enforces the equality of the 5D gauge transformations $R$ and $L$ at the point $z = l_1$. We call $h \left( x \right)$ their common value
\begin{eqnarray}
  h \left( x \right) & \equiv & R \left( x, z = l_1 \right) \hspace{1em} =
  \hspace{1em} L \left( x, z = l_1 \right) . 
\end{eqnarray}
For the vector combination, we adopt a $h$-independent condition
\begin{eqnarray}
  R_{5 \mu} \left( x, z = l_1 \right) + L_{5 \mu} \left( x, z = l_1 \right) &
  = & 0 .  \label{R5+L5IR}
\end{eqnarray}
This simple breaking is enough to give the correct order of magnitude for the
$\chi$PT low-energy constants, $L_i$'s. Essentially, the way it works from the
4D point of view is by lifting the axial tower of resonances from the vector
ones. This will produce at low energies a non-zero decay constant of the pion,
$f$.

\subsubsection{The spectrum}

Given the field content and symmetry-breaking pattern just described, the
mass eigenstates will be the axial and vector combinations of the $L_M$ and
$R_M$ gauge fields, $A_M$ and $V_M$. Two remarks are in order here: the
Lorentz index $M$ takes 5 values, i.e. $A_M = ( A_{\mu}, A_5 )$ and each
component $A_M ( x^{\mu}, z )$ is a linear combination of an infinite number
of 4D fields, $A_M^{( n )} ( x )$, weighted by a profile in the fifth
dimension, $\varphi^A_{ n} ( z )$. This is the so-called KK-decomposition
\begin{eqnarray}
  A_M ( x, z ) = \sum_{n = 0}^{\infty}  & \text{$\varphi^A_{ n} ( z )$}
  A_M^{( n )} ( x ), & 
\end{eqnarray}
where the values of the profiles are completely determined by the geometry and the BCs.

Given our BCs, the physical spectrum will be composed by (i) a multiplet of
massless 4D particle, related to the zero mode of the axial field, the
$A_5^{\left( 0 \right)} ( x )$ and (ii) two sets of 4D massive vector
fields, $A_{\mu}^{( n )}$ and $V_{\mu}^{( n )}$, where $n$ runs from $1$ to
$\infty$. The other KK modes are either absent due to the BCs $\left( \text{no } V_{\mu}^{( 0 )}, A_{\mu}^{( 0 )}, V_{5}^{( 0 )} \right)$, or eaten via higgsing: the $V_5^{( n
)}$ and $A_5^{( n )}$ (with $n \neq 0$) feed the longitudinal components of
the 4D vector fields $V_{\mu}^{( n )}, A_{\mu}^{( n )}$.

After imposing the BCs (\ref{R-LIR}, \ref{R5+L5IR}), part of the full 5D gauge
invariance is broken. The residual symmetries are: (i) chiral
symmetry (or shift symmetry) protecting the  GBs and
(ii) higgsed $\prod_n \bignone [ \tmop{SU} ( N_f ) \times \tmop{SU} ( N_f )
]_{( n )}$ gauge invariance enjoyed by $A_{\mu}^{( n )}$ and $V_{\mu}^{( n
)}$.

\subsubsection{Conformal invariance in the UV}

Consider the two-point correlators $\Pi_{V, A}$ of quark currents $J_{V, A}^{a \mu} = \overline{q} \gamma^{\mu} \left( \gamma_5 \right) \frac{T^a}{2} q$
\begin{eqnarray}
2  \mathi \int \mathd^4 x \mathe^{\mathi q \cdot x} \left\langle 0 \left| T \{ J_{V, A}^{a \mu}
  ( x ) J_{V, A}^{b \nu} ( 0 ) \} \right|0 \right\rangle & = & \delta^{ab}  \left( q^{\mu}
  q^{\nu} - q^2 \eta^{\mu \nu} \right) \Pi_{V, A} \left( q^2 \right) . 
  \label{def2point}
\end{eqnarray}
In large-$N_c$ QCD with vanishing quark masses, the OPE expansion of $\Pi_{V,
A}$ gives, for large euclidean momentum $Q^2 \equiv - q^2$
{\cite{Shifman:1979bx}} \footnote{The hypothesis of
  vacuum
  saturation
 has been used as a simplification  to rewrite the dimension six condensates in
  terms of the quark condensate.}
\begin{eqnarray}
  \Pi_{V, A} \left( - Q^2 \right) & = & - \frac{N_c}{12 \pi^2}  \left\{
  \lambda + \log \left( \frac{Q^2}{\mu^2} \right) +\mathcal{O}( \alpha_s )
  \right\} + \kappa_4  \frac{\alpha_s \langle G G \rangle}{Q^4} + 
  \kappa_6^{V, A}  \frac{\alpha_s \langle \overline{q} q \rangle^2}{Q^6} + \ldots 
  \label{2pointgen}
\end{eqnarray}
where dimensional regularization has introduced a term $\lambda = \mu^{d - 4} 
\frac{2}{d - 4} + \gamma_E - \ln \left( 4 \mathpi \right) + \ln \left( \mu^2
\right) +\mathcal{O} \left( d - 4 \right)$. The first term in
(\ref{2pointgen}) is the dimension-0 operator obtained from perturbative QCD,
the second is a gluon condensate, and the third involves the quark condensate.
For the 5D model to reproduce the partonic logarithm, the metric has to be
conformally/scale invariant near the UV brane: although the result is known in
5D, we will reobtain it explicitly in the computation of the two-point
function, equation (\ref{HEPVA}).

For most of this paper, we adopt the simplest geometry that reproduces the
dimension-0 term in the OPE of $\Pi_{V, A}$. We consider a space-time warped
by the Anti-de-Sitter (AdS) curvature
\begin{eqnarray}
  w \left( z \right) & = & \frac{l_0}{z} .  \label{AdS}
\end{eqnarray}
The metric is then exactly conformal, and conformal invariance is broken only
by the presence of the IR brane, as in the model of Randall and Sundrum
{\cite{Randall:1999ee}}, RS1.

With the warp factor (\ref{AdS}) and our field content, one cannot reproduce
the remaining terms in the OPE of $\Pi_{V, A}$ (\ref{2pointgen}). To remedy
this, deviations from conformality should be introduced near the IR brane: in Section \ref{condensates} we
simply sketch the consequences for the case of the two-point correlators
(\ref{def2point}). With this in mind, we derive results for a generic metric,
except in explicit computations. We also illustrate the dependence on the
metric through numerical comparison between RS1 and flat space (FS).

\section{Chiral implementation of the holographic recipe} \label{field-redef}

Quarks and gluons provide the correct description of QCD at high energies whereas, at
low energies, $\chi$PT is the procedure to compare with experiment. We intend
to link both sides of QCD via a model of GBs and spin-1 mesons, without aiming
for the full generality of an effective theory.

Resemblance with QCD will come from various aspects. The straightforward
reason why a 5D YM model could be of any help to interpolate between QCD and
$\chi$PT is the immediate inclusion of an infinite number of resonances which
respect the correct symmetries (chiral and scale invariance). Other reasons to
look at a 5D model are {\tmem{hints}}: holography and the AdS/CFT
correspondence, in addition to the partial successes of hidden local symmetry
approaches.

In this Section we introduce the so-called \tmem{holographic recipe} derived from the
AdS/CFT correspondence, used to construct the Green's functions of currents. This is
a 5D-specific implementation of the external field method. The external field
method is also often used in descriptions of QCD in terms of mesons
{\cite{Gasser:1984yg}}. It introduces sources to generate the Noether currents
of the 4D $\tmop{SU} \left( N_f \right) \times \tmop{SU} \left( N_f \right)$
global symmetry, see Section \ref{genfunct}. We show how to go from the 5D version to the 4D formalism: in Section
\ref{thea5}, we define pions and vector fields that transform
covariantly with respect to the chiral symmetry.

\subsection{The generating functional} \label{genfunct}

\subsubsection{Green's functions of QCD currents} \label{QCD}

The generating functional of the QCD Noether currents $J_{L, R}^{\mu} = \overline{q} \gamma^{\mu} \frac{1 \mp \gamma_5}{2} \frac{T^a}{2} q$ of the
global $\tmop{SU} \left( N_f \right) \times \tmop{SU} \left( N_f \right)$
symmetry is defined through the functional integral over gluons $G$ and quarks
$q$
\begin{eqnarray}
  \mathcal{Z}_{\tmop{QCD}} [ \ell_{\mu}, r_{\mu} ] & = & \mathe^{\mathi
  \Gamma_{\tmop{QCD}} [ \ell_{\mu}, r_{\mu} ]}  \hspace{1em} = \hspace{1em}
  \int [ \mathd G ] [ \mathd q ] \mathe^{\mathi \int
  (\mathcal{L}_{\tmop{QCD}} + \ell^a_{\mu} J^{a \mu}_L + r^a_{\mu} J^{a \mu}_R
  )} .  \label{ZQCD}
\end{eqnarray}
In the above, $\ell_{\mu} \left( x \right) = \ell^a_{\mu} \frac{T^a}{2}$ and $r_{\mu}
\left( x \right) = r_{\mu}^a \frac{T^a}{2}$ are classical configurations coupled to the
QCD currents.

Green's functions obtained from $\Gamma_{\tmop{QCD}}$ satisfy all Ward
identities of the global 4D $\tmop{SU} \left( N_f \right) \times \tmop{SU}
\left( N_f \right)$ symmetry. The whole set of Ward identities is equivalent
to the fact that $\Gamma_{\tmop{QCD}} [ \ell_{\mu}, r_{\mu} ]$ is invariant
under a {\tmem{local}} version of the 4D $\tmop{SU} \left( N_f \right) \times
\tmop{SU} \left( N_f \right)$ symmetry
{\cite{Gasser:1984yg,Leutwyler:1994iq,Knecht:1995}} which acts on the sources
$\ell_{\mu}, r_{\mu}$ as
\begin{eqnarray}
  \ell_{\mu} & \longmapsto & g_L \ell_{\mu} g_L^{\dag} + \mathi g_L
  \partial_{\mu} g_L^{\dag},  \label{localgL}\\
  r_{\mu} & \longmapsto & g_R r_{\mu} g_R^{\dag} + \mathi g_R \partial_{\mu}
  g_R^{\dag},  \label{localgR}
\end{eqnarray}
i.e. the sources are gauge connections for the local version of chiral
symmetry. In addition, if one coupled QCD to the electroweak interactions, a
subset of these sources would become the dynamical photon and $W^{\pm}, Z^0$
fields: this yields the right interactions of QCD currents with the
electroweak vector bosons.

Consider now another generating functional describing the same Green's
functions in terms of physical fields at low energies, e.g. pions for
$\chi$PT. In order for its Green's functions to obey the same Ward identities,
one requires invariance of the action under the local chiral symmetry, acting on
the mesons {\tmem{and}} the sources. We now turn to our model, and show that
it does indeed satisfy this requirement.

\subsubsection{From 5D to 4D} \label{Zs}

We start
with the 5D action (\ref{action}), integrated over the 5D YM gauge fields
\begin{eqnarray}
  \mathcal{Z}_{\tmop{model}}  & = & \int_{} [ \mathd L_M ] [ \mathd R_M ]
  \mathe^{\mathi \mathcal{S}_{\tmop{YM}} [ L_M, R_M ]} .  \label{3.5}
\end{eqnarray}
The holographic recipe in addition defines that the 5D YM
fields should satisfy the following constraint: their value at the UV boundary
is a classical configuration
\begin{eqnarray}
  L_{\mu} ( x, z = l_0 ) & = & \ell_{\mu} ( x ),  \label{sources}\\
  R_{\mu} ( x, z = l_0 ) & = & r_{\mu} ( x ),  \label{sources2}
\end{eqnarray}
on which the generating functional depends. This holographic recipe will allow
us to extract the Noether currents of the global $\tmop{SU} \left( N_f \right)
\times \tmop{SU} \left( N_f \right)$ symmetry, expressed in terms of mesons.

The generating functional is invariant under the local transformations of
the sources (\ref{localgL}-\ref{localgR}): this 4D local invariance is the 5D
gauge invariance at a specific point $z = l_0$. Therefore, we see hat the
Green's functions will satisfy the QCD Ward identities. Note that electroweak
interactions in this model reside on the UV brane, where the sources are.
Integrating over the boundary values $\ell_{\mu}, r_{\mu}$ would amount to considering dynamical $W^{\pm},
Z^0$ and photon.

After redefining the fields so as to explicitly respect chiral symmetry in
Section \ref{thea5} and extracting the pions, we will perform the KK decomposition. This will leave us with two infinite
towers of massive vector fields $\left( V_{\mu}^{( n )}, A_{\mu}^{( n )}
\right)$, the classical sources $\left( \ell_{\mu}, r_{\mu} \right)$ and the
massless pion modes, collected in the unitary matrix $U $. We will be able to rewrite the generating functional as 
\begin{eqnarray}
  \mathcal{Z}_{\tmop{model}} [ \ell_{\mu}, r_{_{\mu}} ] & = & \int \bignone_{}
  [ \mathd U ] [ \mathd V_{\mu}^{( n )} ] [ \mathd A_{\mu}^{( n )} ] \mathe^{\mathi
  \int d^4 x ( \text{$\mathcal{L}_2^{\chi \tmop{PT}} +\mathcal{L}_4^{\chi
  \tmop{PT}}$} +\mathcal{L}_{\tmop{resonances}} )} ,  \label{Zmodel4D}
\end{eqnarray}
where $(\mathcal{L}_2^{\chi \tmop{PT}} +\mathcal{L}_4^{\chi \tmop{PT}} ) [ U,
\ell_{\mu}, r_{\mu} ]$ are the $\mathcal{O}( p^2 )$ and $\mathcal{O}( p^4 )$
chiral lagrangians. All terms involving resonances are collected in $\mathcal{L}_{\tmop{resonances}} [ U, V_{\mu}^{( n )},
A_{\mu}^{( n )}, \ell_{\mu}, r_{\mu} ]$: kinetic and mass terms, as well as interactions  with pions and sources. At energies
of the order of the resonance masses, the action in (\ref{Zmodel4D}) is the
appropriate one, since it describes the interactions of each resonance
separately. We will obtain and describe it in Section \ref{resonances}. At
energies lower than the KK scale $p^2 \ll \mathpi^2 / l_1^2$, one can
integrate out the massive states: the low-energy action for GBs $\mathcal{L}_2^{\chi \tmop{PT}} +\mathcal{L}_4^{\chi \tmop{PT}} $ is then obtained, with definite predictions for the low-energy constants (Section \ref{GBs}). At the other extreme, i.e. high energies, it is more convenient to use the 5D description, see Section \ref{2point}.

\subsection{The $A_5$ and the pion field} \label{thea5}

We mentioned the spectrum of our model in Section \ref{Zs}. In particular,
contrary to the 5D model by {\cite{Erlich:2005qh,daRold:2005zs}}, we have
introduced only YM fields: the spectrum contains a massless 4D spin-0 mode,
due to the chosen boundary conditions. It is related to the (axial combination
of) the fifth component of the gauge fields. In this section, we show how to
extract the GB field with appropriate symmetry properties.  Our procedure enjoys some similarities
with that of \cite{Son:2003et,Sakai:2004cn}: details are relegated to
Appendix \ref{AA}.

\subsubsection{GBs as the Wilson line}

We first define Wilson lines $\xi$ extending from $l_1$ to
$z$ along the fifth coordinate $x^5$, but for a fixed point $x^{\mu}$ in 4D
space
\begin{eqnarray}
  \xi_R \left( x, z \right) & \equiv & \text{P} \left\{ \mathe^{\mathi
  \int_{l_1}^z \mathd x^5 R_5 \left( x, x^5 \right) \bignone} \right\} ,
\end{eqnarray}
with $\text{P}$ the path-order integral. Using one such Wilson line to go from the UV brane to the IR brane with the
$L_5$ field, and back with the $R_5$ field, we define the 4D object $U$
\begin{eqnarray}
  U ( x ) & \equiv & \xi_R \left( x, l_0 \right) \xi_L^{\dag} \left( x, l_0
  \right).
\end{eqnarray}
Given the properties of the Wilson line, $U$ obeys the following transformation law
under the 5D gauge transformations
\begin{eqnarray}
  U \left( x \right) & \longmapsto & g_R \left( x \right) U \left( x \right)
  g_L^{\dag} \left( x \right) .  \label{U-trsf}
\end{eqnarray}
Equation (\ref{U-trsf}) involves only the elements $\left( g_R, g_L \right)$
of the 4D $\tmop{SU} \left( N_f \right) \times \tmop{SU} \left( N_f \right)$
symmetry. It is the appropriate transformation law for GBs of the breaking to
the vector $\tmop{SU} \left( N_f \right)$ \cite{Coleman:1969sm}. With a vacuum such that $R_5 = 0$ and $L_5 = 0$ up to a gauge, we see that, in our model,
the 4D symmetry is {\tmem{spontaneously}} broken to its vector subgroup.
Indeed, the vacuum is $U = 1$, which is invariant only if $g_R = g_L$.

\subsubsection{Redefinitions for vector fields}

We have extracted the pion field as a product of Wilson lines extending from
one boundary to the other. We also recall the property of the Wilson line
$\xi$ from $l_1$ to a point in the bulk at position $z$
\begin{eqnarray}
  \xi_R^{\dag}  \left( \partial_5 - \mathi R_5 \right) \xi_R & = & 0 . 
  \label{gauge-away}
\end{eqnarray}
This shows that the fifth component of the gauge fields could be gauged away
by using $\xi$ as the gauge transformation. However, such a gauge
transformation depending on the value of fields is more appropriately treated
as a field redefinition, rather than a gauge fixing
{\cite{Gross:1972pv,Grosse-Knetter:1993nn}}. One then defines the following
vector and axial combinations of gauge fields
\begin{eqnarray}
  \hat{V}_M, \hat{A}_M & \equiv & \mathi \left\{ \xi^{\dag}_L  \left(
  \partial_M - \mathi L_M \right) \xi_L \pm \left(L\to R \right) \right\} .  \label{hatted}
\end{eqnarray}
Note that their fifth component vanish, $\hat{V}_5 = 0, \hat{A}_5 = 0$.

The next step is to separate the hatted quantities
into {\tmem{dynamical}} fields and external sources. This is done by
subtracting from the hatted quantities (\ref{hatted}) their values at the
boundary, times the appropriate profile depending on $z$. The UV boundary
value of $\hat{V}_{\mu}$ is still related to the sources, so this subtraction
indeed separates sources and fields. For the vector case, subtracting a
constant profile in $z$ is adequate
\begin{eqnarray}
  V_{\mu} \left( x, z \right) & \equiv & \hat{V}_{\mu} \left( x, z \right) -
  \hat{V}_{\mu} \left( x, l_0 \right) .  \label{Vmu}
\end{eqnarray}
As for the axial combination, its UV boundary value involves the derivative of
the pion field in addition to the axial combination of the sources
\begin{eqnarray}
  \left. \hat{A}_{\mu} \left( x, z \right) \right|_{z = l_0} & = & -
  \frac{\mathi}{2} \xi_R^{\dag} \left( x, l_0 \right)  \left( D_{\mu} U
  \right) \xi_L \left( x, l_0 \right) .  \label{AhatBC}
\end{eqnarray}
In that case, we need to introduce a function $\alpha \left( z \right)$ such
that
\begin{eqnarray}
  A_{\mu} \left( x, z \right) & \equiv & \hat{A}_{\mu} \left( x, z \right) -
  \alpha \left( z \right)  \hat{A}_{\mu} \left( x, l_0 \right) .  \label{Amu}
\end{eqnarray}
The appropriate expression for $\alpha \left( z \right)$ will be provided
later in Section \ref{GBs}. It will follow uniquely from the requirement of
diagonalizing quadratic terms in the action, and therefore turns out to depend
only on the metric.

We have thus constructed fields that transform homogeneously under the adjoint
representation of $h \left( x \right)$
\begin{eqnarray}
  V_{\mu} \left( x, z \right) & \longmapsto & h \left( x \right) V_{\mu}
  \left( x, z \right) h \left( x \right)^{\dag},  \label{Vadjoint}\\
  A_{\mu} \left( x, z \right) & \longmapsto & h \left( x \right) A_{\mu}
  \left( x, z \right) h \left( x \right)^{\dag} .  \label{Aadjoint}
\end{eqnarray}
This implementation of the holographic recipe is very close to the external
field method traditionally used in $\chi$PT {\cite{Gasser:1984yg}}. Another advantage of our procedure is that we
avoid gauge-fixing. This is of paramount importance in order to track down the
chiral symmetry, which severely constrains the interactions of the GBs.

\subsubsection{Derived BCs}

We now derive the BCs that the new fields $V_{\mu}, A_{\mu}$ satisfy. This is
done by injecting the above definitions for the fields $V_{\mu}, A_{\mu}$ into
the BCs for the original fields $R_M, L_M$ on the IR brane (\ref{R-LIR}) and
(\ref{R5+L5IR}). We get {\footnote{This is a
covariant expression, i.e. $\left. \partial_5 V_{\mu}\right|_{l_1} \longmapsto h \left(
\left. \partial_5
V_{\mu} \right|_{l_1}\right) h^{\dag}$.}}
\begin{eqnarray}
  V_{\mu} |_{l_0} & = & 0,  \label{VUV}\\
  \partial_5 V_{\mu} |_{l_1} & = & 0 .  \label{VIR}
\end{eqnarray}
This means that the dynamical field $V_{\mu}$ satisfies what we will call $( -, +
)$ BCs. The UV BC directly follows from having performed the
subtraction (\ref{Vmu}).

Regarding the axial field $A_{\mu}$, we first have to specify BCs for the
function $\alpha \left( z \right)$, which enters into the definition
(\ref{Amu}). We choose the following, which will yield simple BCs for
$A_{\mu}$
\begin{eqnarray}
  \alpha \left( l_0 \right) & = & 1,  \label{alphaUV}\\
  \alpha \left( l_1 \right) & = & 0.  \label{alphaIR}
\end{eqnarray}
Once these BCs for the function $\alpha$ are chosen, it follows that $A_{\mu}$
satisfies $\left( -, - \right)$ BCs, i.e.
\begin{eqnarray}
  A_{\mu} |_{l_0} & = & 0,  \label{AUV}\\
  A_{\mu} |_{l_1} & = & 0 .  \label{AIR}
\end{eqnarray}
Having appropriately subtracted boundary values from the covariant
combinations of $L_{\mu}, R_{\mu}$, we have obtained $V_{\mu},
A_{\mu}$ fields with simple BCs. The sources have been separated, and
will appear explicitly in the lagrangian, as opposed to implicitly
via EOMs in the original action (\ref{3.5}).

\section{Very low energies: Goldstone bosons} \label{GBs}

In this paper, we propose a model of mesons which interpolates between QCD at
low and high energies. We first focus on low energies, where symmetries are
the main handle: the properties of the GBs follow from the (broken) symmetry.
Since the model implements the appropriate spontaneous breaking, we will
recover the non-linear GBs derivative interactions, {\tmem{provided}} our way
of treating the model respects the symmetries. This was the reason behind the
non-standard non-linear treatment of Section \ref{field-redef}.

The $\chi$PT lagrangian is expanded according to powers of four-dimensional
derivatives. Simply counting the number of 4D Lorentz indices in the two terms
of the 5D YM action $F_{\mu 5}^2$ and $F_{\mu \nu}^2$, one can see where the
$\mathcal{O} \left( p^2 \right)$ and $\mathcal{O} \left( p^4 \right)$ terms of
the $\chi$PT lagrangian come from in our model: we write schematically
\begin{eqnarray}
  \left\langle F_{5 \mu} \right\rangle^2 & \rightarrow & \mathcal{L}_2^{\chi
  \tmop{PT}} + \text{resonance mass terms},  \label{F5mu}\\
  \left\langle F_{\mu \nu} \right\rangle^2 & \rightarrow & \mathcal{L}_4^{\chi
  \tmop{PT}} + \text{resonance kinetic \& interaction terms} .  \label{Fmunu}
\end{eqnarray}
Since this is a model with two parameters, we will obtain predictions for the
$\chi$PT low-energy constants. The explicit results are derived in this
Section and  Appendix \ref{AA}. Covariance with respect to 4D chiral symmetry is
automatic, due to the work done in Section \ref{field-redef}, so the
identification with the $\chi$PT lagrangian holds strictly.

As is usual in models where resonances are introduced as fields with one (4D)
Lorentz index as opposed to tensor fields, the contribution of resonance
exchange to GB interactions for $p^2 \ll M_{\rho}^2$ starts at $\mathcal{O}
\left( p^6 \right)$ {\cite{Ecker:1989yg}}. Thanks to the work of the previous Section, ``integrating
out'' resonances is trivial at $\mathcal{O} \left( p^2 \right)$ and
$\mathcal{O} \left( p^4 \right)$: one simply reads off the GB interactions
from the lagrangian $\mathcal{L}_2^{\chi \tmop{PT}} +\mathcal{L}_4^{\chi
\tmop{PT}}$, disregarding operators involving resonance fields.

\subsection{$\chi$PT lagrangian at $\mathcal{O} \left( p^2 \right)$}
\label{Op2}

The rewriting (\ref{F5mu}) is directly obtained from
\begin{eqnarray}
  2 \left\langle L_{5 \mu} L_{5 \nu} + R_{5 \mu} R_{5 \nu} \right\rangle & = &
  \left( \partial_5 \alpha \right)^2  \left\langle D_{\mu} UD_{\nu} U^{\dag}
  \right\rangle + 4 \left\langle \partial_5 V_{\mu} \partial_5 V_{\nu} +
  \partial_5 A_{\mu} \partial_5 A_{\nu} \right\rangle .  \label{R5muR5mu}
\end{eqnarray}
The first term in the right-hand side will yield a kinetic term for the GBs,
once integrated over the fifth coordinate. The second term will yield masses
for the KK modes of $V_{\mu}$ and $A_{\mu}$, after integration by parts.

In fact, there should be an additional term in (\ref{R5muR5mu}), producing a
mixing between the pion and the axial resonances. However, if we ask that
$\alpha$ satisfies the following wave equation
\begin{eqnarray}
  \partial_5 \left( \sqrt{g} g^{\mu \nu} g^{55} \partial_5 \right) \alpha & =
  & 0,  \label{EOMalpha}
\end{eqnarray}
then, the mixing vanishes. We indeed have the freedom to require
(\ref{EOMalpha}), since the function $\alpha$ was up to now completely
undetermined, except for its BCs (\ref{alphaUV}-\ref{alphaIR}). Note that
(\ref{EOMalpha}) is a massless spin-1 equation. This was expected, since
$\alpha ( z )$ is the wave-function of $D_\mu U ( x )$, i.e. of the derivative of a
massless spin-0 field. Bear in mind however, that $\alpha$ has to satisfy
the unusual BCs (\ref{alphaUV}-\ref{alphaIR}).

Our rewritings yield the standard GB interactions involving two derivatives
\begin{eqnarray}
  &  & - \frac{2}{4 g_5^2}  \bigintlim_{l_0}^{l_1} \bignone \mathd z \sqrt{g}
  g^{55} g^{\mu \nu}  \left\langle R_{5 \mu} R_{5 \nu} + L_{5 \mu} L_{5 \nu}
  \right\rangle \nonumber\\
  & = & \frac{f^2}{4}  \left\langle D_{\mu} UD^{\mu} U^{\dag} \right\rangle +
  \frac{1}{2}  \sum_n \left\langle M_{V_n}^2 V^{\left( n \right)}_{\mu} V^{\left( n
  \right) \mu} + M_{A_n}^2 A^{\left( n \right)}_{\mu} A^{\left( n \right) \mu}
  \right\rangle,  \label{5mu5mu}
\end{eqnarray}
with an expression for the pion decay constant $f$
\begin{eqnarray}
  \eta^{\mu \nu} f^2 & \equiv & - \frac{1}{g_5^2}  \bigintlim_{l_0}^{l_1}
  \mathd z \sqrt{g} g^{55} g^{\mu \nu}  \left( \partial_5 \alpha \right)^2 . 
  \label{fpi}
\end{eqnarray}
The first term in the right-hand side of (\ref{5mu5mu}) is the
chiral-invariant $\mathcal{O} \left( p^2 \right)$ kinetic and interaction term
of GBs, including their coupling to sources: it is strictly that of
{\cite{Gasser:1985gg}}. One also obtains a second term in the right-hand side
of (\ref{5mu5mu}), to be interpreted later in Section \ref{KK} as the sum of
the mass terms for the KK modes.

Solving for $\alpha$ (Appendix \ref{AA}), the general expression for $f$ (\ref{fpi}) yields
\begin{eqnarray}
  \left. f^2 \right|_{\text{FS}, \tmop{RS} 1} & = & \frac{1}{\left( g_4 l_1
  \right)^2}, \frac{2}{\left( g_4 l_1 \right)^2},  \label{fpiFS}
\end{eqnarray}
for the flat space and RS1 metrics respectively. In this result, the limit
$l_0 \longrightarrow 0$ has already been taken. For RS1, this limit can be
safely taken only for physical quantities, provided in addition $l_0 / g_5^2$
remains finite. Indeed, one can define the dimensionless couplings
\begin{eqnarray}
  \left. \frac{1}{g_4^2} \right|_{\text{FS}, \tmop{RS} 1} & = &
  \frac{l_1}{g_5^2}, \frac{l_0}{g_5^2},  \label{g4}
\end{eqnarray}
which should be kept finite.

\subsection{Predictions for $\chi$PT lagrangian at $\mathcal{O} \left( p^4
\right)$} \label{0p4}

Particularizing to $N_f=3$, we  turn our attention to the next-to-leading order interactions in the
chiral expansion. Such GB $\mathcal{O} \left( p^4 \right)$ interactions are
produced by those terms in the 5D Yang-Mills lagrangian that contain four 4D
Lorentz indices, i.e. $\left\langle R_{\mu \nu} R_{\rho \sigma} + L_{\mu \nu}
L_{\rho \sigma} \right\rangle$, as described in (\ref{Fmunu}). We provide the
full rewritings in Appendix \ref{AA}: this directly yields interaction terms
of GBs and sources $r_{\mu}, \ell_{\mu}$ respecting chiral symmetry, in the
form of the $\mathcal{O} \left( p^4 \right)$ lagrangian of
{\cite{Gasser:1985gg}}. One then simply identifies the low-energy
constants describing the interactions of GBs at next-to-leading order. Their
values are given by integrals of polynomials of the function $\alpha$ over the
fifth coordinate. We find explicitly
\begin{eqnarray}
  \eta^{\mu \nu} \eta^{\rho \sigma} L_1 & = & \frac{1}{32 g_5^2} 
  \bigintlim_{l_0}^{l_1} \bignone \mathd z \sqrt{g} g^{\mu \rho} g^{\nu
  \sigma}  \left( 1 - \alpha^2 \right)^2,  \label{L1}\\
  \eta^{\mu \nu} \eta^{\rho \sigma} L_{10} & = & - \frac{1}{4 g_5^2} 
  \bigintlim_{l_0}^{l_1} \bignone \mathd z \sqrt{g} g^{\mu \rho} g^{\nu
  \sigma}  \left( 1 - \alpha^2 \right),  \label{L9}\\
 \eta^{\mu \nu} \eta^{\rho \sigma}  H_1 & = & - \frac{1}{8 g_5^2}  \bigintlim_{l_0}^{l_1} \bignone \mathd z
  \sqrt{g} g^{\mu \rho} g^{\nu \sigma}  \left( 1 + \alpha^2 \right) . 
  \label{H1bulk}
\end{eqnarray}
The other $L_i$ constants are given in terms of the previous ones by
\begin{eqnarray}
  L_2 & = & 2 L_1,  \label{L2}\\
  L_3 & = & - 6 L_1,  \label{L3}\\
  L_9 + L_{10} & = & 0 .  \label{L9+L10}
\end{eqnarray}
These relations hold for any metric, and deserve some comments. First of all,
since we have introduced neither scalar sources nor quark masses, the
$\mathcal{O} \left( p^4 \right)$ constants $L_4, L_5, L_6, L_7, L_8$
and $H_2$ do not
appear.

Second, we find the relation $L_2 = 2 L_1$, as expected in large-$N_c$
{\cite{Gasser:1985gg}}. We also find $L_3 = - 6 L_1$, which is standard for
models with one multiplet of vector resonances. It comes directly from the
Skyrme structure of the $\mathcal{O}\left(p^4\right)$ GB interaction term, which we find here as was the case in the extra-dimensional model of
{\cite{Sakai:2004cn}} and hidden local symmetry models
{\cite{Meissner:1988ge}}. Consequences on GB forward scattering amplitudes
will be discussed in Section \ref{Froi}.

Third, we find the relation $L_9 + L_{10} = 0$. This is more unusual, and will
have consequences on the axial form factor in Section \ref{AFF}. In 4D models
with only the $\rho$ resonance (no higher resonances, and no axial
resonances), it was also obtained as a consequence of the KSFR I relation
{\cite{Ecker:1989yg}}. (The result was however modified by the adjunction of
the $a_1$.) In our model, it holds in the presence of both towers of vector and
axial resonances: the two come from linear combinations of the same 5D YM
fields.

 \TABULAR[p]{|c|c|c|}{\hline
 & $    \left( z \right) = 1 $&$ w \left( z \right) = \frac{l_0}{z}$\\
     \hline
  $   f^2  $&$ \frac{1}{\left( g_4 l_1 \right)^2} $&$ \frac{2}{\left( g_4 l_1
     \right)^2}$\\
     \hline
   $  L_1  $&$ \frac{1}{60 g_4^2} $&$ \frac{11}{768 g_4^2}$\\
     \hline
  $   L_2  $&$ 2 L_1 $&$ 2 L_1$\\
     \hline
  $   L_3  $&$ - 6 L_1 $&$ - 6 L_1$\\
     \hline
  $   L_9  $&$ \frac{1}{6 g_4^2} $&$ \frac{3}{16 g_4^2}$\\
     \hline
  $   L_{10}  $&$ - L_9 $&$ - L_9$\\
     \hline}{Formulas for low-energy constants.\label{TfL}}

In Table \ref{TfL}, we give the analytic expressions for the $L_i$'s: we consider two examples for the metric: flat space $w
\left( z \right) = 1$  and RS1 $w \left( z \right) = l_0 / z$ where the limit
$l_0 \longrightarrow 0$ with $g_4$ (\ref{g4}) finite is assumed.

As in any resonance model used at tree-level, the predictions for the
$L_i$'s are scale-independent. Indeed, in the limit of large number of colors
it is expected that the leading order consists of tree diagrams involving
mesons {\cite{Ecker:1989te}} {\footnote{Note that the formalism of
antisymmetric tensor fields is currently being used for treating the lightest
resonances at the level of quantum loops {\cite{Rosell:2004mn}}. This may remedy
the lack of scale dependence of the $L_i$'s obtained by integration of the
resonances at tree-level {\cite{Cata:2001nz,Sanz-Cillero:2002bs}}.}}.
Presumably, the predictions for the $L_i$'s should be interpreted as the
values of the renormalized constants at some scale $\mu$ of the order of the
mass of the first resonances. A comparison of these predictions with the
values extracted  from $\mathcal{O} \left( p^4
\right)$ fits  {\cite{Bijnens:1994qh}}, with the scale $\mu = M_{\rho}$ is presented for our two backgrounds in
Table \ref{Tnum}.
\TABULAR[p]{|c|c|c|c|}{\hline
      &$ w \left( z \right) = 1 $&$ w \left( z \right) = \frac{l_0}{z} $&$
     \text{Experiment}$\\
      \hline
    $ 10^3 L_1 $&$   0.52 $&$ 0.52 $&$ 0.4 \pm 0.3$\\
     \hline
    $ 10^3 L_2 $&$ 1.03 $&$ 1.04 $&$ 1.35 \pm 0.3$\\
     \hline
    $ 10^3 L_3   $&$ - 3.10 $&$ - 3.12 $&$ - 3.5 \pm 1.1$\\
     \hline
    $ 10^3 L_9   $&$ 5.2 $&$ 6.8 $&$ 6.9 \pm 0.7$\\
     \hline
    $ 10^3 L_{10}  $&$ - 5.2 $&$ - 6.8 $&$ - 5.5 \pm 0.7$\\
     \hline}{Numerical values for the $L_i$ coefficients.\label{Tnum}}
 To obtain numerical values, we have to input the two independent parameters of
the model: $g_4$ and $l_1$. For this, we invert the exact formulas yielding
$M_{V_1}$ and $f$, and adjust them to the experimentally measured $M_{\rho}
\simeq 776 \tmop{MeV}$ and $f_{\pi} \simeq 87 \tmop{MeV}$ in the chiral limit.
We stick to this simple procedure throughout the paper since our model is one of
large-$N_c$ QCD, and we therefore do not expect fine numerical agreement (in
particular for resonance masses). As a matter of fact, the only reason why we keep three
digits for $L_1, L_2, L_3$ is not because we trust them, but to show the
difference between the two metrics.

\section{Intermediate energies: spin-1 resonances} \label{resonances}

The KK modes of the 5D model we propose are to be interpreted as mesons: this
is the description appropriate at intermediate energies. The link between QCD
and this model of mesons treated at tree level is provided by the limit of a
large number of colors. In this limit, QCD is equivalent to a theory involving
an infinite number of infinitely narrow states. Our model treats an infinite
tower of states of spin-1, in addition to the GBs. In this Section, properties
and interactions of resonances are extracted.

\subsection{Resonances as Kaluza-Klein excitations} \label{KK}

Following the standard KK reduction procedure, we assume separation of
variables, i.e. that a 5D vector field $V_{\mu} \left( x, z \right)$
satisfying the EOMs derived from the 5D lagrangian can be decomposed as a sum
of 4D modes $V_{\mu}^{\left( n \right)} \left( x \right)$, each possessing a profile
$\varphi_n^V \left( z \right)$ in the fifth dimension
\begin{eqnarray}
  V_{\mu} \left( x, z \right) & = & \sum_n V^{\left( n \right)}_{\mu} \left( x
  \right) \varphi^V_n \left( z \right) . 
\end{eqnarray}
From the EOMs, we find that the profiles $\varphi_n^V$ have to satisfy the
following differential equation for a Sturm-Liouville operator
\begin{eqnarray}
  \eta^{\mu \nu} \partial_5 \left( \sqrt{g} g^{\rho \sigma} g^{55} \partial_5
  \right) \varphi^V_n & = & \sqrt{g} g^{\mu \nu} g^{\rho \sigma} M_{V_n}^2
  \varphi^V_n,  \label{EOM}
\end{eqnarray}
with $\left( -, + \right)$ BCs. $M_{V_n}^2 > 0$ will be identified as the mass
of the 4D KK mode $V^{\left( n \right)}$. The same procedure can be applied to
the $A_{\mu} \left( x, z \right)$ field, with $\left(-,-\right)$ BCs. Explicit formulas for wave-functions
are collected in Appendix \ref{AB} for the case of flat space and RS1.
Approximate expressions for the masses of the heavy resonances $M_{V_n, A_n},
n \gg 1$ are given in Table \ref{TMn}.
The indexing is such that $\rho$
and $a_1$ are labeled by $n = 1$, i.e. correspond to the KK modes $V^{\left( 1
\right)}$ and $A^{\left( 1 \right)}$ respectively. These expressions are
approximate at a few percent level for $n \sim 1$ in the case of RS1, and
exact for flat space. The real value of $M_{V_1}$ in RS1 is $2.4 / l_1$. For heavy excitations $n \gg 1$, the
behavior for these 5D models is
\begin{eqnarray}
  M_{V_n, A_n}^2 & \propto & n^2 . 
\end{eqnarray}
\TABULAR[p]{|c|c|c|}{\hline
       &$ w \left( z \right) = 1 $&$ w \left( z \right) = \frac{l_0}{z}$\\
     \hline
$     M_{V_n}   $&$ \frac{\mathpi}{l_1}  \left( n - \frac{1}{2} \right) $&$
     \frac{\mathpi}{l_1}  \left( n - \frac{1}{4} \right)$\\
     \hline
 $    M_{A_n}   $&$ \frac{\mathpi}{l_1} n $&$ \frac{\mathpi}{l_1}  \left( n +
     \frac{1}{4} \right)$\\
     \hline}{Approximate expressions for the masses of heavy resonances.\label{TMn}}
This behavior of the masses with $n$ is a consequence of the 5D Lorentz
invariance broken by the finite size of the extra-dimension. This may seem
unsatisfactory, given the known result for large-$N_c$ QCD in 1+1 dimensions
{\cite{'tHooft:1974hx}} and the Regge behavior {\cite{Anisovich:2000kx}}.
Nonetheless, there exists no proof that $M_n^2 \sim n$ is the right behavior for $n \gg 1$
in large-$N_c$ QCD in 4D.

Another objection that may be put forward regards the spectrum of the light
resonances. In RS1 with the limit $l_0 \rightarrow 0$, the ratio $M_{A_1} /
M_{V_1}$ is approximately 1.6, in agreement with the experimental value of $M_{a_1}
/ M_{\rho}$. However, in our model, the quick growth of the interspacing
between resonances as $n$ increases is already felt for the lightest modes,
implying that the masses $M_{V_2}, M_{V_3}$ are higher than the experimental
values quoted by {\cite{PDG}} for $M_{\rho'}, M_{\rho''}$, see Table \ref{TMrho}. In flat space, the
problem would be worse.
We note that this can in principle be cured directly
by including modifications of the AdS$_5$ metric near the IR brane. At any
rate, the leading order in $1 / N_c$ falls short of taking into account the
non-negligible width of QCD resonances.

\TABULAR[h]{|c|c|c|c|}{\hline
     $ w \left( z \right) = 1 $&$ w \left( z \right) = \frac{l_0}{z} $&$
     \text{Experiment}$\\
         \hline
    $ M_{A_1} =  1.6 \tmop{GeV} $&$ M_{A_1} = 1.2 \tmop{GeV} $&$
	 M_{a_1} = 1230 \pm 40 \tmop{MeV}$\\
     \hline
    $ M_{V_2} = 2.3 \tmop{GeV} $&$  M_{V_2} =  1.8 \tmop{GeV} $&$
     M_{\rho'} =  1465 \pm 25 \tmop{MeV}$\\
     \hline
     $M_{V_3} = 3.9 \tmop{GeV} $&$  M_{V_3} = 2.8 \tmop{GeV} $&$ M_{\rho''} = 1688.1 \pm 2.1 \tmop{MeV}$\\
     \hline}{Numerical values for lightest resonance
       masses\label{TMrho}.}

\subsection{Couplings of resonances}

The original 5D YM lagrangian contains interactions between 5D fields. These
entail, via the KK decomposition (\ref{KK}), definite resonance interactions.
In Appendix \ref{AC}, we give expressions for couplings involving one
resonance as bulk integrals of polynomials in $\alpha$ times the appropriate
factors of the wave-functions.

\TABULAR[p]{|c|c|c|}{\hline
     &$ w \left( z \right) = 1 $&$ w \left( z \right) = \frac{l_0}{z}$\\
     \hline
    $ f_{V_n} $&$ \frac{\sqrt{2}}{\mathpi g_4}  \frac{1}{n - \frac{1}{2}} $&$
     \frac{1}{g_4}  \frac{1}{\sqrt{n - \frac{1}{4}}}$\\
     \hline
    $ f_{A_n} $&$ \frac{\sqrt{2}}{\mathpi g_4}  \frac{1}{n} $&$ \frac{1}{g_4} 
     \frac{1}{\sqrt{n + \frac{1}{4}}}$\\
     \hline
    $ g_{V_n} $&$ \frac{\sqrt{2}}{\mathpi^3 g_4}  \frac{1}{\left( n - \frac{1}{2}
     \right)^3} $&$ \frac{\left( - 1 \right)^n}{8 \sqrt{2} \mathpi^2 g_4} 
     \frac{1}{\left( n - \frac{1}{4} \right)^2}$\\
     \hline}{Approximate formulas for resonance parameters.\label{Tn}}

Using the notations of {\cite{Prades:1993ys}},
generalized to resonances indexed by $n$, the expressions of Table \ref{Tn}
show the behavior of the couplings $f_{V_n}, f_{A_n}, g_{V_n}$ for large $n$ \footnote{The
constants $f_{V_n}$, defined via $\left. \left\langle 0| J_V^{\mu} \left( 0 \right)
\right| V^{\left( n \right)} \left( \varepsilon, p \right) \right\rangle =
\varepsilon^{\mu} f_{V_n} M_{V_n}^2 $ describe the decay of the mesons into
leptons when the electroweak sector is coupled to the QCD currents. $g_{V_n}$ describes the coupling of vector mesons to two pions.}.
Once again, these expressions are approximate in the case of RS1, and exact
for flat space.
For RS1, the approximation is again good up to a few percents, even for $n =
1$, except for $g_{V_n}$, for which the few-percent level accuracy is reached
only for $n > 15$. Indeed, even the sign of $g_{V_1}$ would be wrong if one
used the asymptotic formula.

Table \ref{Gammarho} collects the results for $\rho$ decays. Inputs are $f
= 87$ MeV and $M_{\rho} = 776$ MeV. The first column corresponds to
a model with only one vector resonance: the $\rho$, after imposing the KSFR I
\& II relations $g_{\rho}^2 M_{\rho}^2 = f^2 / 2$ and$f_{\rho} = 2 g_{\rho}$
to fix $g_{\rho}$ and $f_{\rho}$. The 5D RS1 model is certainly not disfavored
compared to the KSFR predictions in the chiral limit.

\TABULAR[h]{|c|c|c|c|c|}{\hline
&$ \text{KSFR} $&$ w \left( z \right) = 1 $&$ w \left( z \right) =
     \frac{l_0}{z} $&$ \text{Experiment}$\\
     \hline
$     \Gamma \left( \rho \rightarrow e^+ e^- \right) \tmop{KeV} $&$ 4.4 $&$ 4.4 $&$    8.1 $&$ 7.02 \pm 0.11$\\
     \hline
$     \Gamma \left( \rho \rightarrow \pi \pi \right) \tmop{MeV} $&$ 205 $&$ 135 $&$   135 $&$ 150.3 \pm 1.6$\\
     \hline}{$\rho$ decay widths\label{Gammarho}.}

We also note the following relation between couplings of the axial resonances
defined in the lagrangian (\ref{5.10}) of Appendix \ref{AC},
\begin{eqnarray}
  \alpha_{A_n} & = & - \frac{1}{2 \sqrt{2}} f_{A_n} .  \label{alphaA}
\end{eqnarray}
Relation (\ref{alphaA}) implies the vanishing of the coupling of any axial
resonance $A^{\left( n \right)}$ to one pion and a vector source/photon
{\cite{Knecht:2001xc}}. The decay $a_1 \to \pi \gamma$ does not
occur at this order in the minimal model. (The same is true for
all axial resonances.) Higher-derivative terms would modify
this result, but should only be considered at higher orders. In Section \ref{condensates}, we will
introduce modifications of the metric that are felt differently by the axial
and vector combinations of gauge fields (in order to account for local
order-parameters of chiral symmetry). We plan to investigate how this modifies
$\Gamma \left(a_1 \to \pi \gamma\right)$ in a forthcoming paper.

To illustrate how the currents (quark bilinears in QCD) are expressed in terms of mesons, we use the expressions provided in the Appendix \ref{AC}. One can then derive a
field-current identity \cite{Sakurai:1969}
\begin{eqnarray}
  J^{a}_{V \mu} ( x ) & =& \sum^{\infty}_{n = 1}  f_{V n} M_{V n}^2 V^{a ( n )}_{ \mu} ( x ) + \text{non-linear terms} ,  \label{field-current}
\end{eqnarray}
with the expressions for the decay constants given by {\footnote{Such a result
can also be understood from a more holographic point of view, by plugging in
the KK decomposition in equation (6) of {\cite{Erlich:2005qh}}.}}
\begin{eqnarray}
  \eta^{\mu \nu} f_{V_n} M^2_{V_n} & = & \left. - \frac{\sqrt{2}}{g_5^2} 
  \sqrt{g} g^{\mu \nu} g^{55} \partial_5 \varphi^V_n \right|_{l_0} . 
\end{eqnarray}
Similarly, we get the corresponding formula for the axial current
\begin{eqnarray}
  J^a_{A \mu} ( x ) &= & f \partial_{\mu} \pi^a + \sum^{\infty}_{n = 1}  f_{A
  n} M_{A_n}^2 A^{a ( n )}_{\mu} ( x ) + \text{non-linear terms} .
\end{eqnarray}

\section{High energies: sums of resonance contributions} \label{HE}

In this Section, we examine the following amplitudes: $\left\langle \pi \left|
J_V^{\mu} \right| \pi \right\rangle$, $\left\langle \gamma \left| J_A^{\mu}
\right| \pi \right\rangle$. The Lorentz-invariant functions involved are
called vector form factor and axial form factor. We also consider
forward elastic scattering of GBs $\left\langle \pi \pi | \pi \nonesep \pi
\right\rangle$. We discuss how many subtractions are needed for the
corresponding dispersion relations to converge.

Our interest in these questions is motivated by the precedent of 4D models
with a finite number of resonance multiplets. In this case, one imposed that
vector form factor, axial form factor and GB forward-scattering amplitudes satisfy dispersion relations
with an appropriate number of subtraction (respectively 0, 0 and 1). This
 soft high-energy behavior is the one expected from QCD {\cite{Ecker:1989yg}}. This also brought the
different formalisms for massive spin-1 fields into agreement.

\subsection{Sums over resonance couplings}
\label{SR}

Before proceeding, we make the following remark, of use for the present
Section. In our 5D model, we automatically obtain the following sum rules
\begin{eqnarray}
  \sum_{n = 1}^{\infty} f_{V_n} g_{V_n} & = & 2 L_9,  \label{fVgV}\\
  \sum_{n = 1}^{\infty} f_{V_n} g_{V_n} M_{V_n}^2 & = & f^2, 
  \label{fVgVMV2}\\
  \sum_{n = 1}^{\infty} g_{V_n}^2 & = & 8 L_1,  \label{L1SR}\\
  \sum_{n = 1}^{\infty} g_{V_n}^2 M_{V_n}^2 & = & \frac{f^2}{3} . 
  \label{KSFRsum}
\end{eqnarray}
They are derived using the completeness relation for the KK wave-functions.
One starts from the explicit expression for the resonance couplings of
Appendix \ref{AC}, and uses as well the EOMs and BCs for the various
wave-functions. The sum rules are thus a consequence of the gauge symmetry and
the 5D locality. The expressions (\ref{fpi}, \ref{L1}, \ref{L9}) for the
low-energy constants $f, L_1, L_9$ appearing in the right-hand side of
(\ref{fVgV}-\ref{KSFRsum}) are finite for the metrics of interest. Convergence
can be checked explicitly for flat space and RS1, using the behavior with $n$
of couplings and masses, Tables \ref{TMn} and \ref{Tn}.

The relations (\ref{fVgV}-\ref{L1SR}) will look familiar from 4D models of resonances: in that case,
one had finite sums rather than infinite ones. One required relations such as
(\ref{fVgV}-\ref{L1SR}) in order to guarantee the QCD behavior at high energies \cite{Ecker:1989yg}.
(Certain 4D models provide some of these relations automatically.) In this Section, we explore the consequences of
(\ref{fVgV}-\ref{L1SR}) on the high-energy behavior of some amplitudes.

\subsection{Vector form factor}

The vector form factor $F \left( q^2 \right)$ is defined as $\left(q=p-p'\right)$
\begin{eqnarray}
  \left\langle \pi^a \left( p' \right) \left| J_V^{b \mu} \left( 0 \right)
  \right| \pi^c \left( p \right) \right\rangle & = & \mathi \varepsilon^{a b
  c}  \left( p+p' \right)^\mu F \left( q^2 \right) .
\end{eqnarray}

\subsubsection{4D point of view}

In this Section, we describe the vector form factor in terms of exchange of (an infinite
number) of resonances. From a partonic analysis
{\cite{Lepage:1979zb,Lepage:1980fj}}, it is expected that the vector form factor satisfies an
unsubtracted dispersion relation. We show that such a soft behavior is
satisfied in our model.

The vector form factor gets contributions from the $\mathcal{O}( p^2 )$ term in the chiral
lagrangian (\ref{5mu5mu}), the $L_9$ term in the $\mathcal{O}( p^4 )$
Lagrangian, and the exchange of resonances
\begin{eqnarray}
  F \left( q^2 \right) & = & 1 + 2 L_9  \frac{q^2}{f^2} + \frac{q^4}{f^2} 
  \sum_{n = 1}^{\infty} \frac{f_{V_n} g_{V_n}}{M_{V_n}^2 - q^2} .  \label{Ft}
\end{eqnarray}
This expression is convenient for low energies. The high-energy behavior of the
infinite sum (\ref{Ft}) however, is not completely transparent. This is
because it takes the form of a twice-subtracted dispersion relation for $F
\left( q^2 \right)$. Indeed, reading off the imaginary part of $F \, \left(
q^2 \right)$ from (\ref{Ft})
\begin{eqnarray}
  \frac{1}{\mathpi} \tmop{Im} F ( s ) & = & \frac{1}{f^2}  \sum_{n =
  1}^{\infty} f_{V_n} g_{V_n} M_{V_n}^4 \delta \left( s - M_{V_n}^2 \right), 
  \label{ImF}
\end{eqnarray}
(\ref{Ft}) is equivalent to $F ( q^2 ) = F ( 0 ) + F' ( 0 ) q^2 +
\frac{q^4}{\mathpi}  \bigintlim_0^{\infty} \frac{\mathd s}{s^2} 
\frac{\tmop{Im} F \left( s \right)}{s - q^2}$. As is usual in large-$N_c$
approximations, the dispersive integral has converted to a discrete sum in
(\ref{Ft}), which converges as long as $\sum_n f_{V_n} g_{V_n} / M_{V_n}^2$
converges. However, we can do better than that: since the sum $\sum_n f_{V_n}
g_{V_n} M_{V_n}^2$ converges (Table \ref{Tn}), one finds that the two
subtractions are unnecessary. Using the values for the sums $\sum_n f_{V_n}
g_{V_n} = 2 L_9$ (\ref{fVgV}) and$\sum_n f_{V_n} g_{V_n} M_{V_n}^2 = f^2$
(\ref{fVgVMV2}), one obtains
\begin{eqnarray}
  F ( q^2 ) & = & \frac{1}{f^2}  \sum_{n=1}^\infty f_{V_n} g_{V_n} 
  \frac{M_{V_n}^4}{M_{V_n}^2 - q^2} , \label{FtHE}
\end{eqnarray}
which is an unsubtracted dispersion relation (c.f. the imaginary part (\ref{ImF})).

We emphasize that the result obtained above, i.e. that the vector form factor satisfies an
unsubtracted dispersion relation, does not depend on the metric. Similar
relations are not automatic in 4D models for resonances: even in hidden local symmetry models,
it needs to be imposed by fixing one free parameter. There is therefore a non-trivial
step in the high-energy behavior in going from hidden local symmetry to a 5D model. The possibility of
 rewriting the vector form factor as a sum of poles without contact terms is
indeed specific of a continuous extra-dimension, as the case of a discrete
extra-dimension {\cite{Son:2003et}} demonstrated. The unsubtracted dispersion relation then
follows.

\subsubsection{5D point of view}

Using the 5D language, one obtains from the lagrangian
\begin{eqnarray}
  f^2 F \left( q^2 \right) & = & f^2 + \frac{q^2}{2 g_5^2} 
  \bigintlim_{l_0}^{l_1} \mathd zw \left( z \right) \Gamma ( z ) +
  \frac{q^4}{2 g_5^4} \bigintlim^{l_1}_{l_0} \mathd zw \left( z \right) \Gamma
  ( z ) \bigintlim^{l_1}_{l_0} \mathd z' w \left( z' \right) G_q ( z, z' ), 
  \label{VFF1}
\end{eqnarray}
with $\Gamma \left( z \right) \equiv 1 - \alpha \left( z \right)^2$. This expression is the 5D version of (\ref{Ft}). The 5D
propagator occurring above is the mixed position-momentum propagator $G_q
\left( z, z' \right)$ defined as
\begin{eqnarray}
  G ( x, z ; x', z' ) & = & \int \frac{\mathd^4 p}{( 2 \mathpi )^4}
  \mathe^{\mathi q \cdot ( x - x' )} G_q \left( z, z' \right) . 
\end{eqnarray}
It satisfies
\begin{eqnarray}
  \left\{ q^2 + \frac{1}{w \left( z \right)} \partial_z \left( w \left( z
  \right) \partial_z \right) \right\} G_q ( z, z' ) & = & - \frac{g_5^2 }{w
  \left( z \right)} \delta ( z - z' ) .  \label{propEOMgen}
\end{eqnarray}
The relation with expression (\ref{Ft}) is provided by the KK decomposition of
the 5D propagator

\begin{eqnarray}
  G_q ( z, z' ) & = & \sum_{n = 1}^{\infty}  \frac{\varphi_n^V ( z )
  \varphi_n^V ( z' )}{M_{V_n}^2 - q^2},  \label{GKK}
\end{eqnarray}
and the expressions for the resonance couplings $f_{V_n}$, $g_{V_n}$ of
Appendix \ref{AC}. In 5D language, one can obtain a more compact expression
for the vector form factor by using the EOM for the 5D propagator (\ref{propEOMgen}). We get
\begin{eqnarray}
  f^2 F \left( q^2 \right) & = & \left. f^2 + \frac{q^2}{2 g_5^4} 
  \bigintlim_{l_0}^{l_1} \mathd zw \left( z \right) \Gamma ( z ) \partial_{z'}
  G_q ( z, z' ) \right|_{z' = l_0},  \label{VFF2}
\end{eqnarray}
where the pure $L_9$ term has disappeared. We have used the normalization $w \left( z' = l_0 \right) = 1$.
Finally, one can use again the EOM to absorb the contact term and obtain a
compact form
\begin{eqnarray}
  f^2 F \left( q^2 \right) & = & \left. \frac{1}{2 g_5^4} 
  \bigintlim_{l_0}^{l_1} \mathd zw \left( z \right)  \left\{ \partial_z \Gamma
  ( z ) \right\} \partial_z \partial_{z'} G_q ( z, z' ) \right|_{z' = l_0} . 
  \label{VFF3}
\end{eqnarray}
This expression coincides with the 4D one (\ref{FtHE}).

During this section and the next one, we will use schematic drawings to
represent this kind of relationships, replacing for instance $\int \mathd z' w
\left( z' \right) \bignone$ by $\int' \bignone$, and indicating the 5D
propagator by a double line, and sources by crosses. The three contributions
of equation (\ref{VFF1}) are shown schematically in the first line of
(\ref{FVFF}), and the compact result (\ref{VFF3}) is depicted in the last line
of (\ref{FVFF})
\begin{eqnarray}
  &  & \parbox{1.5cm}{\text{\epsfig{file=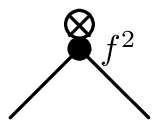}}}
 \hspace{2em} + \hspace{2em} q^2 
  \parbox{1.5cm}{\text{\epsfig{file=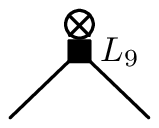}}} \hspace{2em} +
  \hspace{2em} \frac{q^4}{g_5^4} \int \Gamma \int'
  \parbox{1.5cm}{\text{\epsfig{file=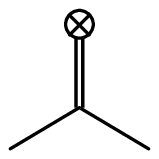}}}  \nonumber\\
  & = & \parbox{1.5cm}{\text{\epsfig{file=tex-figs-interpol-pspage1.eps}}}
  \hspace{2em} + \hspace{2em} \frac{q^2}{g_5^4}  \int \bignone \Gamma
  \partial' \parbox{1.5cm}{\text{\epsfig{file=tex-figs-interpol-pspage3.eps}}}
  \nonumber\\
  & = & \frac{1}{g_5^4}  \int \bignone \left( \partial \Gamma \right)  \partial \partial'
  \parbox{1.5cm}{\text{\epsfig{file=tex-figs-interpol-pspage3.eps}}} . \label{FVFF}
\end{eqnarray}

\subsection{Froissart bound}
\label{Froi}
Having discussed in detail the unsubtracted dispersion relation for the vector form factor, we briefly give
the result for GB scattering. The reasoning parallels that of
{\cite{Ecker:1989yg}}, since we have the relations $L_2 = - 2 L_1$ (\ref{L2})
and $L_3 = - 6 L_1$ (\ref{L3}) {\footnote{These relations were derived
assuming $N_f = 3$: we stick to this case here.}}. We also have the
relation $\sum_n g_{V_n}^2 = 8 L_1$ (\ref{L1SR}), which generalizes the usual
one $g_{\rho}^2 = 8 L_1$ to an infinite sum.

One considers forward $\left( t = 0 \right)$ elastic GB scattering, which
receives contributions from the $\chi$PT lagrangian, through the constants
$L_1, L_2, L_3$, and from resonance exchange. Combining left- and right-handed
cuts using the variable $\nu \equiv \left( s - u \right) / 2$, and using the
relations just mentioned, one shows that the scattering amplitude satisfies a
once-subtracted dispersion relation, as expected from the Froissart bound
{\cite{Froissart:1961ux}}. Indeed, the elastic forward scattering amplitude can be rewritten as
\begin{eqnarray}
  T_{i j} \left( \nu \right) & = & T_{i j} ( 0 ) + \kappa_{i j} 
  \frac{\nu^2}{f^4}  \sum_{n = 1}^{\infty} g_{V_n}^2 
  \frac{M_{V_n}^4}{M_{V_n}^4 - \nu^2},  \label{7.12}
\end{eqnarray}
where $i, j$ are $\tmop{SU} \left( N_f = 3 \right)$ quantum numbers and
$\kappa_{i j}$ is a constant depending on the process. The subtraction
constant $T_{i j} \left( 0 \right)$ vanishes for GBs, from Adler's theorem.

\subsection{Axial form factor}
\label{AFF}

The definition  of the axial form factor $G_A \left( q^2 \right)$ is $\left( q=p-p'\right)$
\begin{eqnarray}
  \left\langle \gamma \left( \epsilon, p' \right) \left| J_A^{a \mu} \left( 0
  \right) \right| \pi^b \left( p \right) \right\rangle & = & \sqrt{2} \frac{e}{f} \left( p'^\mu p\cdot \epsilon -\epsilon^\mu p\cdot p'\right) G_A \left(
  q^2 \right) + \text{pion pole term}. 
\end{eqnarray}
Our model yields the following 
generalization of the result obtained in {\cite{Knecht:2001xc}}
\begin{eqnarray}
  G_A \left( q^2 \right) & = & 4 \left( L_9 + L_{10} \right) + q^2  \sum_{n =
  1}^{\infty} \frac{f_{A_n}  \left( f_{A_n} + 2 \sqrt{2} \alpha_{A_n}
  \right)}{M_{A_n}^2 - q^2} . \label{aff}
\end{eqnarray}
Plugging in the relations that hold in our model
(independently of the metric), i.e. $L_9 + L_{10} = 0$ (\ref{L9+L10}) and
$f_{A_n} = - 2 \sqrt{2} \alpha_{A_n}$ (\ref{alphaA}), we find that the axial form factor  vanishes identically
\begin{eqnarray}
  G_A \left( q^2 \right) & = & 0 .
\end{eqnarray}
Including higher-derivative terms in the action would produce a non-vanishing
resonance-exchange contribution. This is however of higher-order in the
presumed derivative expansion. In our 5D model, the contact term $4 \left( L_9 +L_{10} \right)$ in (\ref{aff}) vanishes,
which is the only way that $G_A \left( q^2 \right)$ can satisfy an
unsubtracted dispersion relation, given the vanishing of resonance-exchange terms. Therefore,
the axial form factor in our 5D model satisfies an unsubtracted dispersion relation, albeit in a trivial way.
Modifications of the metric in the IR, if they are felt differently by vector and
axial combinations as introduced later in Section
\ref{condensates}, will change this. We plan to investigate this in future work.

\subsection{Lightest vector-meson dominance} \label{LMD}

For completeness, we discuss vector-meson dominance and KSFR I \& II
relations. The fact that the KSFR relations are not obeyed in the model leaves
room for better experimental agreement, see Table \ref{Gammarho}.

The sum rules (\ref{fVgV}-\ref{L1SR}) constitute a generalization of the
relations considered in {\cite{Ecker:1989yg}} in the sense that, on the
resonance (i.e. left-hand) side of equations, the sum involves an infinite
number of terms. In the 5D model, the sum rules hold for any metric, but the
relative contribution of each resonance $V_1, V_2, V_3, \cdots$ depends
on the metric. We ask to what extent the sums are satisfactorily reproduced by
taking into account only the $\rho$, i.e. to which extent the lightest meson
dominates the sum, or the statement of lightest vector-meson dominance (LMD),
see Table \ref{TLMD}. We see that the sums indeed receive a large contribution
from the $\rho$, for both metrics \footnote{This is true for sums involving vector resonances only. On the other hand, saturation by $\rho$ and $a_1$ does not hold for sums such as $\sum_n \left(f_{V_n}^2 M_{V_n}^{2 k}-f_{A_n}^2 M_{A_n}^{2 k}\right)$, i.e. moments of the left-right correlator such as $L_{10}$ and the Weinberg sum rules, see Appendix \ref{AD}.}.

Related to the last line in Table \ref{TLMD} is the question of the KSFR II
relation. Within our notations, the KSFR II relation would be the statement
that the definition
\TABULAR[p]{|c|c|c|}{\hline
 &$ w \left( z \right) = 1 $&$ w \left( z \right) = \frac{l_0}{z}$\\
     \hline
 $    \frac{f_{V_1} g_{V_1}}{\sum_n f_{V_n} g_{V_n}} $&$ 0.99 $&$ 1.02$\\
     \hline
  $   \frac{f_{V_1} g_{V_1} M_{V_1}^2}{\sum_n f_{V_n} g_{V_n} M_{V_n}^2} $&$ 0.81
     $&$ 1.11$\\
     \hline
   $  \frac{g^2_{V_1}}{\sum_n g^2_{V_n}} $&$ 0.999 $&$ 0.999$\\
     \hline
    $ \frac{g^2_{V_1} M_{V_1}^2}{\sum_n g^2_{V_n} M_{V_n}^2} $&$ 0.99 $&$ 0.99$\\
     \hline}{Relative contributions of the $\rho$ resonance
as compared to the total sums of Section \ref{SR}.\label{TLMD}}
\begin{eqnarray}
  \frac{1}{a} & \equiv & \frac{g_{V_1}^2 M_{V_1}^2}{f^2},  \label{KSFRII}
\end{eqnarray}
would yield $a = 2$. In the 5D model, one sees from the sum rule $\sum_n
g_{V_n}^2 M_{V_n}^2 = f^2 / 3$ (\ref{KSFRsum}) that the natural value for the
KSFR II ratio is $3$ rather than $2$. Indeed, the sum (\ref{KSFRsum}) is
dominated by the contribution from the $\rho$ (Table \ref{TLMD}), and the KSFR
II ratio takes a value close to 3, as already noted in {\cite{daRold:2005zs}}.
Regarding the KSFR I relation $f_{\rho} = 2 g_{\rho}$, the only related sum
rule we find in our model is $\sum_{n = 1}^{\infty} \left( 2 f_{V_n} g_{V_n} -
f_{V_n}^2 + f_{A_n}^2 \right) = 4 \left(L_9 + L_{10}\right) = 0$: the axial resonance
contribution cannot be separated. There is therefore no favored value for the
KSFR I ratio: it will depend on the metric.

\section{High energies: two-point functions and OPE} \label{2point}

In this section we compute the two-point functions $\Pi_{V, A}$ in the RS1
case. The RS1 metric indeed has the right behavior near the UV brane to yield
the partonic logarithm in the OPE of $\Pi_{V, A}$ (\ref{2pointgen}). This is discussed below.

We will first derive compact expressions for $\Pi_{V, A}$ following the rules
of 5D. We recall that the 5D treatment preserves the full symmetry structure
of the model and is the suitable one for high energies. In Appendix \ref{AD},
we discuss sum rules related to $\Pi_{L R}$ from the resonance/4D point of
view: the infinite sums over resonances need to be regularized. The two
pictures are related by writing the 5D propagator $G_q \left( z, z' \right)$
as a sum of resonance poles (\ref{GKK}).

\subsection{The vector and axial two-point functions} \label{s6.1}

The vector two-point function is obtained by taking two derivatives with
respect to sources in the generating functional. This leads to contact
contributions coming from the $\mathcal{O} \left( p^4 \right)$ lagrangian and
resonance exchange terms, which can be expressed directly in terms of the 5D
propagator
\begin{eqnarray}
  \Pi_V ( q^2 ) & = & - 4 L_{10} - 8 H_1 + 2 \frac{q^2}{g_5^4}
  \bigintlim^{l_1}_{l_0} \bignone \mathd zw \left( z \right) 
  \bigintlim^{l_1}_{l_0} \bignone \bignone \bignone \mathd z' w \left( z'
  \right) G_q ( z, z' ) .  \label{6.1}
\end{eqnarray}
We can use the equation of motion of the 5D propagator (\ref{propEOMgen}), and
the expressions of $L_{10}$ (\ref{L9}) and $H_1$ (\ref{H1bulk}) to rewrite the
two-point function as
\begin{eqnarray}
  \Pi_V ( q^2 ) & = & \left. \frac{2}{q^2 g_5^4} \partial_z \partial_{z'} G_q
  ( z, z' ) \right|_{z = z' = l_0} .  \label{PIV1}
\end{eqnarray}
In the above, we have assumed the normalization $w \left( z = l_0 \right) = 1$
for the warp factor at the UV brane. Going from equation (\ref{6.1}) to
(\ref{PIV1}) is schematized as
\begin{eqnarray}
  \Pi_V & = & \text{\epsfig{file=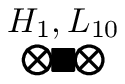}}
  \hspace{2em} + \hspace{2em} \frac{q^2}{g_{5^{}}^4} \int \int'
  \text{\epsfig{file=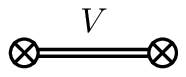}} \nonumber\\
  & = & \left. \frac{1}{q^2 g_5^4} \partial \partial'
  \text{\epsfig{file=tex-figs-interpol-pspage7.eps}} \right|_{l_0,
  l_0} . 
\end{eqnarray}
In the first expression, the contact terms appear explicitly outside the
integral involving the 5D resonance field propagator. In the final expression,
only the second derivative of the 5D propagator evaluated at the boundary
appears. This embodies the idea of holography: one could have used a completely equivalent way of obtaining the two-point function:
by using directly the field-current identity (\ref{field-current}) in the
expression for the two-point function (\ref{def2point}). This directly 
yields the ``holographic-looking'' result (\ref{PIV1}).

The same remarks hold for the axial case, in which case there are
contributions from the $\mathcal{O} \left( p^2 \right)$ lagrangian
(\ref{5mu5mu}) via a pion pole and a contact term, yielding a transverse
structure. We write, with the pion propagator represented by a simple line
\begin{eqnarray}
  \Pi_A & = & \text{\epsfig{file=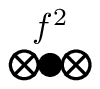}}
  \hspace{2em} + \hspace{2em}
  \text{\epsfig{file=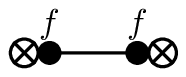}} \hspace{2em} +
  \hspace{2em} \text{\epsfig{file=tex-figs-interpol-pspage5.eps}}
  \hspace{2em} + \hspace{2em} \frac{q^2}{g_5^4}  \int \int' \alpha
  \text{\epsfig{file=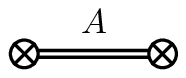}} \alpha
  \nonumber\\
  & = & \left. \frac{1}{q^2 g_5^4} \partial \partial'
  \text{\epsfig{file=tex-figs-interpol-pspage8.eps}} \right|_{l_0,
  l_0} . 
\end{eqnarray}
As depicted above, the compact expression obtained after using twice the EOM
for the 5D propagator (\ref{propEOMgen}) includes the pion pole term. This is
due to the BCs for $\alpha$ and for $A_{\mu}$.

\subsubsection{Explicit expression for RS1}

In this Section, we specialize to the case of RS1. The propagator in RS1
satisfies the EOM (\ref{propEOMgen}) with $w \left( z \right) = l_0 / z$, i.e.
\begin{eqnarray}
  \left\{ q^2 + z \partial_z \left( \frac{1}{z} \partial_z \right) \right\}
  G_q ( z, z' ) & = & - g_4^2 z \delta ( z - z' ) .  \label{propEOM}
\end{eqnarray}
This leads to the explicit form involving Bessel functions
\begin{eqnarray}
  G_q ( z, z' ) & = & - \frac{g_4^2 z z'}{\alpha - \beta} \{ I_1 ( Q z_> ) -
  \alpha K_1 ( Q z_> ) \} \{ I_1 ( Q z_< ) - \beta K_1 ( Q z_< ) \}, 
\end{eqnarray}
where $z_{<, >} = \min ( \max ) ( z, z' )$. Depending on the boundary
conditions ($+$ or $-$), the values of $\alpha$ and $\beta$ are given by
\begin{eqnarray}
  \text{BC } + / - \text{ at } l_1 : \alpha_{\pm} & = & \frac{- I_0 ( Q l_1
  )}{K_0 ( Q l_1 )}, \frac{I_1 ( Q l_1 )}{K_1 ( Q l_1 )},  \label{alpha}\\
  \text{BC } + / - \text{ at } l_0 : \beta_{\pm} & = & \frac{- I_0 ( Q l_0
  )}{K_0 ( Q l_0 )}, \frac{I_1 ( Q l_0 )}{K_1 ( Q l_0 )} .  \label{beta}
\end{eqnarray}
After simplifications, the two-point functions $\Pi_{V, A}$ exhibit a quite
compact expressions
\begin{eqnarray}
  \Pi_{V, A} \left( - Q^2 \right) & = & \frac{2}{g_4^2}  \frac{\left\{ I_0 ( Q
  l_0 ) + \alpha_{\pm} K_0 ( Q l_0 ) \right\}  \left\{ I_0 ( Q l_0 ) + \beta_-
  K_0 ( Q l_0 ) \right\} }{\alpha_{\pm} - \beta_-} . 
\end{eqnarray}
We now assume that the momentum involved remains below $1 / l_0$, and perform
an expansion in terms of $Q l_0$,
\begin{eqnarray}
  \Pi_{V, A} ( - Q^2 ) & = & \frac{- 1}{g_4^2}  \left( \log \frac{Q^2}{\mu^2}
  + \lambda_{\tmop{UV}} ( \mu ) \right) + \frac{2}{g_4^2 \alpha_{\pm} ( Q l_1
  )} +\mathcal{O}( Q^2 l_0^2 ),  \label{HEPVA}
\end{eqnarray}
where $\lambda_{\tmop{UV}} ( \mu ) = \log ( \mu^2 l_0^2 / 4 ) + 2 \gamma_E $. We have introduced a renormalization scale $\mu$, independent of the regulator $1 /
l_0$. The expression (\ref{HEPVA}) is valid for all energies below the
regulator $Q \ll 1 / l_0$, whether $Q$ is below $M_{\rho} \sim 2.4 / l_1$ or
above it.

The $\Pi_{V, A}$ describe Green's functions of operators which, in QCD, are
composite (quark bilinears). Nonetheless, one can send the regulator $l_0
\longrightarrow 0$ and still get finite results for $\Pi_{V, A}$, provided a
counter-term $\delta \left( z - l_0 \right)  \bignone H^{\text{brane}}_1 
\left\langle R_{\mu \nu} R_{\mu \nu} + L_{\mu \nu} L_{\mu \nu} \right\rangle$
is added to the 5D YM lagrangian (\ref{action}). Such a term is allowed by the
symmetries of the model and yields, after the field redefinitions of Section
\ref{field-redef}, an additional contribution to the source contact term $H_1$
(\ref{H1bulk}). If the coefficient $H_1^{\text{brane}}$ depends on $l_0$ so as
to compensate the divergence of (\ref{HEPVA}) when $l_0 \longrightarrow 0$,
this allows us to remove the regulator $l_0 \longrightarrow 0$ in
(\ref{HEPVA}), and consider arbitrarily large momenta. In the following we
always assume that momenta are smaller than the regulator $1 / l_0$, or that
the regulator is sent to infinity.

Comparing with the QCD OPE expansion (\ref{2pointgen}), we identify the
combination of parameters of the 5D RS1 model that correspond to $N_c$ \cite{Son:2003et}
\begin{eqnarray}
  N_c & = & \frac{12 \mathpi^2}{g_4^2} .  \label{Nc}
\end{eqnarray}
We can now go back to the expressions of Section \ref{GBs} and
\ref{resonances} and check the large-$N_c$ scaling of quantities like the pion
decay constant, the $L_i$ constants, as well as the resonance masses and
couplings. We find as expected from \cite{'tHooft:1974jz,Witten:1979kh,Gasser:1985gg}
\begin{eqnarray}
  f^2, L_1, L_2, L_3, L_9, L_{10}, f_{V_n}^2, g_{V_n}^2, f_{A_n}^2 & \propto & N_c,
  \\
  M_{V_n}, M_{A_n} & \propto & N_c^0 . 
\end{eqnarray}
Using as inputs $f_{\pi} = 87$ MeV and $M_{\rho} = 776$ MeV, we find $N_c \simeq
4.3$ instead of 3. This is the reason why we have used $M_{\rho}$ rather than $N_c$ as an
input to give numerical values for low-energy data. The order of magnitude for
$N_c$ is right, but the situation may still be improved by taking into account
deformations of the metric, as we suggest in Section \ref{condensates}. This
would take into account condensates, not limited to $\left\langle \overline{q} q
\right\rangle$ and without introducing the additional degrees of freedom of
{\cite{Erlich:2005qh,daRold:2005zs}}.

The second piece in the correlator (\ref{HEPVA}) is the spontaneous symmetry
breaking produced by different BCs in the IR
\begin{eqnarray}
\text{for } Q \gg 1/l_1, \hspace{2em}  \frac{1}{\alpha_{\pm}} & \simeq & \mp \mathpi e^{- 2 Q l_1} .  \label{falloff}
\end{eqnarray}
From the 5D point of view the interpretation of this term is clear: it is
caused by the breaking of conformal invariance due to the finite size of a compact
extra-dimension. This breaking will be manifest only at energies below the
compactification scale, the scale of the first resonance
{\cite{Randall:1998uk}}. The fact that vector and axial fields feel different
BCs in the IR will produce a nonzero value of the left-right correlator at low energies but
exponentially vanishing at high ernergies. Interestingly, (\ref{falloff})  adopts the form of a
violation of the quark-hadron duality by a finite size instanton
{\cite{Shifman:2000jv}}.

To see how the two-point function $\Pi_V$ contains already an infinite number
of resonances, let us write it for time-like momenta $q^2 > 0$
\begin{eqnarray}
  g_4^2 \Pi_V ( q^2 ) & = & - \lambda_{\tmop{UV}} \left( \mu \right) - \log
  \left( \frac{q^2}{\mu^2} \right) + \mathpi \frac{Y_0 ( ql_1 )}{J_0 ( ql_1 )}
  .  \label{PIV2}
\end{eqnarray}
Taking the high-energy limit $ql_1 \gg 1$, one sees explicitly the two
contributions: the logarithm and the sum over an infinite number of
resonances
\begin{eqnarray}
  \text{for } q \gg 1 / l_1, \hspace{2em} g_4^2 \Pi_V ( q^2 ) & \simeq & -
  \lambda_{\tmop{UV}} \left( \mu \right) - \log \left( \frac{q^2}{\mu^2}
  \right) + \tan \left( ql_1 - \frac{\mathpi}{4} \right) . 
\end{eqnarray}
At low energies $ql_1 \ll 1$, the logarithmic contribution is exactly canceled
by part of the second term in (\ref{PIV2}) to end up with
\begin{eqnarray}
  \text{for } q \ll 1 / l_1, \hspace{2em} g_4^2 \Pi_V ( q^2 ) & \simeq & 2
  \log \left( \frac{l_1}{l_0} \right) + \frac{1}{2} ( q l_1 )^2 + \frac{5}{64}
  ( q l_1 )^4 + \mathcal{O}\left(q^6 l_1^6\right). 
\end{eqnarray}
Note that the constant term is the expected one since, in RS1, we have $2 \log
\left( l_1 / l_0 \right) = - 4 L_{10} - 8 H_1$. Again, the limit $l_0
\longrightarrow 0$ can be taken if the additional contribution
$H_1^{\text{brane}} \left( l_0 \right)$ is considered.

\subsection{The left-right correlator}

From the result for $\Pi_{V, A}$  (\ref{HEPVA}), we see that the left-right
correlator in this model is described in terms of Bessel functions, that are
valid for any energy below the UV regulator $1 / l_0$
\begin{eqnarray}
  \Pi_{L R} ( - Q^2 ) & \equiv & \frac{1}{2}  \left( \Pi_V \left( - Q^2 \right) -
  \Pi_A \left( - Q^2 \right) \right) \hspace{1em} = \hspace{1em} \frac{-
  1}{g_4^2 Q l_1 }  \frac{1}{I_0 ( Q l_1 ) I_1 ( Q l_1 )} +\mathcal{O} \left(
  Ql_0 \right) . 
\end{eqnarray}
We can explore two regimes, low and high energies, depending on how the
momentum is compared with the resonance mass scale, $1 / l_1$. For large
euclidean momentum
\begin{eqnarray}
  \text{for } Q \gg 1 / l_1, \hspace{2em} \Pi_{L R} (- Q^2 ) & \simeq & \frac{-
  2}{g_4^2} \mathpi \mathe^{- 2 Q l_1} .  \label{exp}
\end{eqnarray}
For large $Q$, the left-right correlation function vanishes faster than any
power of $Q^2$
\begin{eqnarray}
  \lim_{Q^2 \rightarrow +\infty} Q^{2 n} \Pi_{L R} (- Q^2 ) & = & 0,
  \hspace{2em} \text{for all } n .  \label{noLOP}
\end{eqnarray}
This means that no local order parameter appears in its OPE expansion. In
particular, the dimension three condensate $\left\langle \overline{q} q
\right\rangle$ is absent {\cite{Hirn:2004ze}}. The exponential
fall-off at high energies {\cite{Barbieri:2003pr}} is readily understood in
the model, since the symmetry is broken on the IR brane, at a finite distance
from us in the fifth dimension, and therefore generates only non-local order
parameters. We will see how this translates in terms of resonance parameters
in Appendix \ref{AD}, yielding Weinberg sum rules.

For very low momentum, $Q \ll 1 / l_1$,
\begin{eqnarray}
 \text{for }Q \ll 1/l_1,  \hspace{2em} \Pi_{L R} ( - Q^2 ) & \simeq & \frac{1}{g_4^2}  \left( - \frac{2}{\left( Ql_1
  \right)^2} + \frac{3}{4} - \frac{17}{96}  \left( Ql_1 \right)^2 \right)
  +\mathcal{O} \left(  Q^4 l_1^4 \right) . 
\end{eqnarray}
This expression takes the standard form expected from $\chi$PT with large-$N_c$ : $\Pi_{L R} ( Q^2 ) = - f_{\pi}^2 / Q^2 - 4 L_{10}
+\mathcal{O}( Q^2 )$: we recover the values for $f $ and $L_{10}$ obtained in
(\ref{fpiFS}) and Table \ref{TfL}.

\subsection{4D point of view} \label{2point4D}

The above results for the two-point correlators were derived directly using
the 5D propagator. It is a general fact in our model that such compact
expressions can be derived in the 5D picture. On the other hand, the imaginary
part of the Green's function involves sums over KK modes, hiding the 5D origin
of the model: in any model with narrow resonances, the imaginary part of
$\Pi_V \left( s \right)$ is given by
\begin{eqnarray}
  \frac{1}{\mathpi} \tmop{Im} \Pi_V \left( s \right) & = & 2 \sum_{n =
  1}^{\infty} f_{V_n}^2 M_{V_n}^2 \delta \left( s - M_{V_n}^2 \right) . 
  \label{ImPiVKK}
\end{eqnarray}
Indeed, the 4D version of equation (\ref{6.1}) is
\begin{eqnarray}
  \Pi_V \left( q^2 \right) & = & - 4 L_{10} - 8 H_1 - q^2  \sum_{n =
  1}^{\infty} \frac{f_{V_n}^2}{M_{V_n}^2 - q^2} .  \label{6.26}
\end{eqnarray}
The equivalence between the two writings (\ref{6.26}) and (\ref{6.1}) is
provided by the relation between 5D and 4D propagators, equation (\ref{GKK}).
Notice that (\ref{6.26}) can be seen as a once-subtracted dispersion relation
obtained from (\ref{ImPiVKK}).

Surely, an expression such as (\ref{ImPiVKK}) cannot reproduce the partonic result
that $\tmop{Im} \Pi_V$ tends to a constant for large time-like momenta
$\frac{1}{\mathpi} \tmop{Im} \Pi^{\text{QCD}}_V \left( s \right) \underset{s
\rightarrow \infty}{\longrightarrow} \frac{N_c}{12 \mathpi^2}$. However, a sum
over an infinite number of terms can reproduce the right logarithmic behavior
for large euclidean momenta $Q^2 \longrightarrow + \infty$, provided the decay
constants for the heavy resonances follow the power-law $f_{V_n}^2 \underset{n
\rightarrow \infty}{\propto} \frac{1}{n}$
{\cite{Beane:2001uj,Golterman:2002mi}}. A look at Table \ref{Tn} shows that
the RS1 metric yields the right behavior in the limit $l_0 \longrightarrow 0$.
Note that one can derive the sum rule
\begin{eqnarray}
 \sum_{n = 1}^{\infty} f_{V_n}^2 & = &  \frac{1}{g_5^2}  \bigintlim_{l_0}^{l_1} \mathd zw \left( z \right) ,  \label{logdiv}
\end{eqnarray}
using the completeness relation as before. For RS1, we have $\frac{w(z)}{g_5^2}=\frac{1}{g_4^2 z}$: the divergence at the UV boundary of the integral in (\ref{logdiv})
when  $l_0 \longrightarrow 0$ is related to the $1 / n$ behavior for the $f_{V_n}^2$. For $l_0 > 0$, the integral (\ref{logdiv}) does
not diverge. In this case, we have seen in (\ref{HEPVA}) that the logarithmic
behavior was reproduced for $Q \ll 1 / l_0$, with corrections analytic in
$Q l_0$.

We can now turn back to our once-subtracted dispersion relation for $\Pi_V$ (\ref{6.26}). We
understand that the subtraction is necessary in the limit $l_0 \longrightarrow
0$ for RS1. However, we have seen that the 5D expression corresponding to this
once-subtracted dispersion relation (\ref{6.1}) could be recast in a more compact form
(\ref{PIV1}). This is so because the 5D description is not encumbered by the
infinite sums which may diverge. In 5D, one gets analytic expressions:
discussing the number of subtractions is then replaced by considering the
behavior at large momenta, which can be done directly as in (\ref{HEPVA}) for
euclidean momenta, or as (\ref{PIV2}) for $q^2 > 0$. This is to be contrasted
with the 4D infinite sums, for which the limit $q^2 \longrightarrow + \infty$
is hampered by the presence of an infinite number of resonances with
masses $M^2 > q^2$ for any given $q^2$. From the 5D expressions, the limit
$\left| q^2 \right| \longrightarrow \infty$ is easily obtained: provided one
stays below $1 / l_0$ or sends it to infinity, the only scale to compare the
momentum to is $1 / l_1$ which fixes {\tmem{all}} resonance masses at once.

\subsection{Condensates} \label{condensates}

Given the breaking by BCs and the exact conformality of the metric, broken
only by the IR brane, the model we have presented here has no local order
parameters. However, any metric that is asymptotically AdS near the UV brane
\begin{eqnarray}
  w ( z ) & \underset{z \rightarrow l_0}{\sim} & \frac{l_0}{z}, 
\end{eqnarray}
will produce the dimension zero term (\ref{2pointgen}) in the OPE of $\Pi_{V,
A}$. 
More generally, one can introduce deformations of conformality in the
bulk, which will reproduce the condensates. In
\cite{Erlich:2005qh,daRold:2005zs}, deformations were introduced by a
5D scalar field obtaining a vacuum expectation value. Here, we wish to
keep the field content minimal: we introduce deformations of the
metric. As was the case for the BCs, the deformations should be felt differently by the axial and vector combinations of fields
\begin{eqnarray}
  w_{V, A} ( z )^2   & =&\left( \frac{l_0}{z} \right)^2  \left( 1 + \sum_{d = 2}
  c_{2 d}^{V, A}  \left( \frac{z - l_0^{}}{l_1 - l_0} \right)^{2 d} \right) .
\end{eqnarray}
These coefficients $c_{2 d}$ parametrize our ignorance
of the dynamics responsible for generating the condensates. They could come
from several interlaced sources in 5D: unknown string dynamics coupling differently to axial and vector fields, back-reaction
of the metric, bulk fields frozen at energies below $l_0$, higher order corrections of the YM action,$\ldots$
Building a model for these modifications is out of the scope of this paper,
which looks for simplicity (in order to catch the origin of the behavior of 5D
models) and aims to interpolate with the known large-$N_c$ QCD behavior, rather than to {\tmem{solve}} QCD.

At linear order in the $c_{2 d}$, the modification in the two point function is
given by {\footnote{We thank Arcadi Santamaria for invaluable
help on this point. }}
\begin{eqnarray}
  \Pi_{V, A} ( Q ) & \simeq & \Pi_{V, A}^0 ( Q ) + \sum_{d = 2} \frac{\langle
  \mathcal{O}_{2 d} \rangle_{V, A}}{Q^{2 d}}, 
\end{eqnarray}
where $\langle \mathcal{O}_{2 d} \rangle_{V, A} = \frac{N_c}{24
  \pi^2} \sqrt{\pi} d^2 \frac{\Gamma \left( d \right)^3}{\Gamma \left( d +
\frac{1}{2} \right)} \frac{c_{2 d}^{V, A}}{l_1^{2 d}}  $. The first term $\Pi_{V,A}^0 ( Q )$ was already given
in (\ref{HEPVA}) and contains the zero-order condensate and the exponential
fall-off (\ref{falloff}). The scale of the condensate is tied with the IR
scale, $l_1$, and will be responsible for the breaking of the generalized
Weinberg sum rules (\ref{falloff}, \ref{GWSRs}) down to the usual two Weinberg
Sum rules.

Note at this point that our aim is not to adjust the coefficients
$c_d^{V, A}$ to the OPE one by one, but to extract information on these high
energy parameters by looking at their subsequent effects on the mismatch with
$N_c = 3$, resonance spectrum, $\Gamma \left( a_1 \longrightarrow \pi \gamma
\right)$, axial form factor$\ldots$ and in turn extract some of
the condensates.

\section{Conclusions}

We have presented a model of GBs and an infinite number of spin-1 resonances.
It is based on a setup with $\tmop{SU} ( N_f ) \times \tmop{SU} ( N_f )$ YM
fields living in five dimensions (5D), with appropriate boundary conditions. This provides a description of
four-dimensional (4D) mesons: massless pions and vector and axial resonances. The 5D
gauge symmetry and conformality of the geometry along the extra dimension
provide the correct symmetries to link with QCD, interpolating between Chiral
Perturbation Theory and large-$N_c$ QCD. The model in its simplest form
contains two parameters.

Note that we have treated gravity as non-dynamical. Its only effect is to provide a particular metric, which determines the gauge field profiles and the condensates. Wether such a metric can be obtained by solving a dynamical higher-dimensional setup or it is just mimicking effects of quark confinement is out of the scope of this paper.

The simplicity and computability of the two-parameter model is manifest when
one compares the inputs (the above-mentioned 5D symmetries) with the outputs.
In particular, we have explored all the energy ranges where the model is
computable: (i) at low energies, we have described the pion interactions in
terms the low energy constants ($L_i$'s) of the chiral lagrangian, (ii) we
have described resonance interactions at intermediate energies and (iii) we
have obtained the two-point functions and form factors as compact expressions
involving the 5D propagator; these can be computed in the high-energy limit,
as well as the low-energy one. Finally, (iv) we have described the relation
between 4D and 5D descriptions. We now summarize the results point by point.

(i) {\tmem{Low energies}}. We have used field redefinitions in order to explicitly exhibit the low-energy properties.
This applies in particular to the symmetries and interactions of the pions. We
have shown that, even in such a constrained setup, one can obtain low energy
constants in agreement with experiment. Regarding the Green's functions, we
have obtained compact expressions involving the 5D propagator. These match the
low-energy chiral expansion, in which non-local order parameters appear.

(ii) {\tmem{Resonances}}. From a 4D point of view, the model describes an
infinite number of narrow spin-1 resonances. A crucial point is that these
descend from a 5D Yang-Mills lagrangian. Therefore, their couplings are given
by overlap integrals of their wave-functions in the fifth dimension. This
means that all resonance couplings (a) are uniquely given as a function of the
gauge coupling once the metric is known, and (b) satisfy sum rules reflecting
the 5D origin of the model. We find that dominance by the lightest vector meson ($\rho$) is automatic.

(iii) {\tmem{High energies}}. We were also able to compare the behavior of the
Green's functions at large euclidean momenta with the operator product
expansion of QCD. In its simpler version, the model lacks the ingredient of
non-vanishing local order parameters of QCD. This is remedied without introducing new degrees of freedom, as described in Section
\ref{condensates}.

We have shown that the model yields an infinite number of Weinberg sum rules
(and only two when the condensates are introduced), that extended vector meson
dominance was automatic, and that the vector form factor followed an
unsubtracted dispersion relation.

(iv) {\tmem{4D and 5D point of view}}. In the 5D model, moments of the
left-right two point correlators are well-defined. The Weinberg sum rules are
obeyed, and generalizations thereof. The 5D result is in itself all we need,
but we have also shown how to interpret it in the Kaluza-Klein picture. We
have discussed the prescription provided by the extra-dimensional setup for
handling the divergent sums over the infinite number of resonances, which has
always been an issue for 4D models. Divergences cancel between contributions
from the tower of vector resonances and its axial counterpart, because both
descend from 5D fields living in the {\tmem{same}} extra-dimension.

\acknowledgments
We benefited from lengthy discussion with Toni Pich, Jorge Portol\'es,
Ximo Prades, Nuria Rius, Arcadi Santamaria, Juanjo Sanz-Cillero and
Yael Shadmi. We also thank them for a critical reading of a draft of
this paper, and Juanjo Sanz-Cillero for providing us with a draft of
his paper \cite{Sanz-Cillero:2005ef}. VS thanks Ami Katz for sharing his results (during the workshop {\tmem{The weak scale, the Planck scale and what's in-between}}  organized by Yael Shadmi and supported by The Institute of Theoretical Physics, Technion), as well as Alex Pomarol. JH is grateful to Marc Knecht and Jan Stern for advice and for their opinion about the draft, and is very happy to thank Bruno Julia-D\'\i az for his continuous encouragement to work on the form factor.
JH is supported by the European network EURIDICE, contract HPRN-CT-2002-00311 and by the Generalitat Valenciana, contract GV05/015.
\begin{appendix}

\section{Field redefinitions} \label{AA}

In this Appendix we show how to go from the model in terms of the original 5D
fields in (\ref{3.5}), $L_M, R_M$ to the 4D fields in (\ref{Zmodel4D}), $U,
A_{\mu}^{( n )}, V_{\mu}^{( n )}, \ell_{\mu} \tmop{and} r_{\mu}$. These last
fields separate clearly between sources and fields, and show that indeed the
pion enjoys the correct transformation properties.

The main property of the Wilson line is
\begin{eqnarray}
  \xi_R \left( x, z \right) & \longmapsto & R \left( x, z \right) \xi_R \left(
  x, z \right) h \left( x \right)^{\dag} .  \label{transf-Gamma}
\end{eqnarray}
Its covariant derivatives are defined as
\begin{eqnarray}
  D_{\mu} \xi_R \left( x, z \right) & \equiv & \partial_{\mu} \xi_R \left( x,
  z \right) - \mathi R_{\mu} \left( x, z \right) \xi_R \left( x, z \right) +
  \mathi \xi_R \left( x, z \right) h_{\mu} \left( x \right),  \label{Dmuw}\\
  D_5 \xi_R \left( x, z \right) & \equiv & \partial_5 \xi_R \left( x, z
  \right) - \mathi R_5 \left( x, z \right) \xi_R \left( x, z \right) . 
\end{eqnarray}
Similar definitions apply to the case of the $L_M$ gauge field as well.
$h_{\mu}$ is the common value of $L_{\mu}$ and $R_{\mu}$ at $l_1$.

Of particular interest is the Wilson line $\xi$ extending from one boundary to
the other $\xi_R \left( x, l_0 \right)$. Other 4D fields that are useful for
computations can be constructed. The first is denoted by $u_{\mu}$, and
contains the {\tmem{non-linear}} derivative of the pion field. This is a more
fundamental object than $U$ since GB interactions can only appear with powers
of momenta
\begin{eqnarray}
  u_{\mu} \left( x \right) & \equiv & \mathi \left\{ \xi_R^{\dag}\left( x,l_0\right) D_{\mu}
  \xi_R\left( x,l_0\right) - \xi_L^{\dag}\left( x,l_0\right) D_{\mu} \xi_L\left( x,l_0\right) \right\} \\
  & = & \mathi \left\{ \xi_R^{\dag}\left( x,l_0\right)  \left( \partial_{\mu} - \mathi r_{\mu}
  \right) \xi_R\left( x,l_0\right) - \xi_L^{\dag}\left( x,l_0\right)  \left( \partial_{\mu} - \mathi \ell_{\mu}
  \right) \xi_L\left( x,l_0\right) \right\} . 
\end{eqnarray}
In the above, the terms involving the $h_{\mu}$ connection in the covariant
derivative (\ref{Dmuw}) cancel out, and we have
\begin{eqnarray}
  u_{\mu} & \longmapsto & hu_{\mu} h^{\dag} . 
\end{eqnarray}
Also of use is the orthogonal combination, which transforms as a connection
for the 4D transformation $h \left( x \right)$
\begin{eqnarray}
  \Gamma_{\mu} \left( x \right) & \equiv & \frac{1}{2}  \left\{ \xi_R^{\dag}\left( x,l_0\right)
  \left( \partial_{\mu} - \mathi r_{\mu} \right) \xi_R\left( x,l_0\right) + \xi^{\dag}_L\left( x,l_0\right)  \left(
  \partial_{\mu} - \mathi \ell_{\mu} \right) \xi_L\left( x,l_0\right) \right\},  \label{Gammamu}\\
  \Gamma_{\mu} & \longmapsto & h \Gamma_{\mu} h^{\dag} + \mathi h
  \partial_{\mu} h^{\dag} . 
\end{eqnarray}

The result $\hat{V}_5=\hat{A}_5=0$ in equation (\ref{hatted}) may seem surprising. Since the fifth component of the gauge fields contain the GB fields,
we cannot get rid of them in the bulk without conflicting with the BCs
{\cite{Contino:2003ve,Sakai:2004cn}}: the pion is a physical field and cannot
be eliminated. There is no contradiction here, since we have defined
{\tmem{new}} fields $\hat{V}_M, \hat{A}_M$, whose BCs still have to be derived
from those imposed on the original fields $R_M, L_M$. In particular, we find
that the pion field has not disappeared, since we have the boundary condition
(\ref{AhatBC}) for $\hat{A}_\mu$, and hence the redefinition (\ref{Amu})  to obtain
$A_\mu$ {\footnote{Comparing with equation (48) of {\cite{Ecker:1989yg}}, we see that
their constant $c$ has turned into a function depending only on the additional
coordinate $z$.}}. Note also that the expression of
$\hat{V}_{\mu}$ on the UV brane involves the sources, through (\ref{Vmu})
{\footnote{The equation corresponds to the one used to obtain
(43) in {\cite{Ecker:1989yg}}.}} where $\hat{V}_{\mu} \left( x, l_0 \right) = \mathi
\Gamma_{\mu} \left( x \right) .$

At this stage, we have used field redefinitions eliminating the 5D fields
$R_M, L_M$ in favor of the $z$-dependent fields $V_{\mu},A_{\mu}$
and the 4D field $u_{\mu}$. In particular, no fifth component of any
vector field will occur in the lagrangian. To see how the rewritings lead to the $\mathcal{O}( p^2 )$
lagrangian (\ref{R5muR5mu}), we focus on the term that
involves indices 5 in the YM action (\ref{action}). We shall need the
identities
\begin{eqnarray}
  L_{5 \mu} & = & \xi_L  \left( \partial_5 V_{\mu} + \partial_5 A_{\mu} -
  \frac{1}{2}  \left( \partial_5 \alpha \right) u_{\mu} \right) \xi_L^{\dag},
  \\
  R_{5 \mu} & = & \xi_R  \left( \partial_5 V_{\mu} - \partial_5 A_{\mu} +
  \frac{1}{2}  \left( \partial_5 \alpha \right) u_{\mu} \right) \xi_R^{\dag}, 
\end{eqnarray}
which follow from using the redefinitions (\ref{Vmu}, \ref{Amu}). $\alpha ( z
)$ is a solution of a massless EOM (\ref{EOMalpha}). For flat space and RS1
respectively, the solution of the EOM for $\alpha$ derived in
(\ref{EOMalpha}), and satisfying the BCs (\ref{alphaUV}-\ref{alphaIR}) is,
respectively
\begin{eqnarray}
  \alpha \left( z \right) |_{\tmop{FS}} & = & 1 - \frac{z}{l_1}, \\
  \left. \alpha \left( z \right) \right|_{\tmop{RS} 1} & = & 1 - \frac{z^2 -
  l_0^2}{l_1^2 - l_0^2} .  \label{alphaRS1}
\end{eqnarray}

The $\mathcal{O}( p^4 )$ lagrangian is obtained from the terms with two 4D
indices in the YM action
\begin{eqnarray}
  \left\langle R_{\mu \nu} R_{\rho \sigma} +
  L_{\mu \nu} L_{\rho \sigma} \right\rangle
  & = & \frac{1}{2}  \left\langle F_{+ \mu \nu} F_{+ \rho \sigma} + F_{- \mu
  \nu} F_{- \rho \sigma} \right\rangle,  \label{RR+LL}
\end{eqnarray}
where the new field strengths $F_{\pm \mu \nu}$ are defined as
\begin{eqnarray*}
  F_{\pm \mu \nu} \left( x, z \right) & \equiv & \xi_L^{\dag} L_{\mu \nu}
  \xi_L \pm \xi_R^{\dag} R_{\mu \nu} \xi_R .
\end{eqnarray*}
We shall also need the definitions of the field strengths of the sources
\begin{eqnarray}
  f_{\pm \mu \nu} \left( x \right) & \equiv & \xi_L^{\dag}\left(x,l_0\right) \ell_{\mu \nu}
  \xi_L\left(x,l_0\right) \pm \xi_R^{\dag}\left(x,l_0\right) r_{\mu \nu} \xi_R\left(x,l_0\right) . 
\end{eqnarray}
One can show that $F_{+ \mu \nu}$ contains the field-strength of the vector
field, the field-strength of the vector source $f_{+ \mu \nu}$ and terms
involving GBs, as follows
\begin{eqnarray}
  F_{+ \mu \nu} & = & 2 \left( \nabla_{\mu} V_{\nu} - \nabla_{\nu} V_{\mu} -
  \mathi \left[ V_{\mu}, V_{\nu} \right] - \mathi \left[ A_{\mu}, A_{\nu}
  \right] \right) \nonumber\\
  & + & \mathi \alpha \left( \left[ u_{\mu}, A_{\nu} \right] + \left[
  A_{\mu}, u_{\nu} \right] \right) + f_{+ \mu \nu} + \mathi \frac{1 -
  \alpha^2}{2}  \left[ u_{\mu}, u_{\nu} \right] .  \label{F+}
\end{eqnarray}
We also obtain the following identity for $F_{- \mu \nu}$, in terms of the
field strength of the axial field, of the field strength of the axial source
$f_{- \mu \nu}$, and additional terms involving GBs
\begin{eqnarray}
  F_{- \mu \nu} & = & 2 \left( \nabla_{\mu} A_{\nu} - \nabla_{\nu} A_{\mu} -
  \mathi \left[ V_{\mu}, A_{\nu} \right] - \mathi \left[ A_{\mu}, V_{\nu}
  \right] \right) \nonumber\\
  & + & \mathi \alpha \left( \left[ u_{\mu}, V_{\nu} \right] + \left[
  V_{\mu}, u_{\nu} \right] \right) + \alpha f_{- \mu \nu} .  \label{F-}
\end{eqnarray}
In the above, we have used the standard definition for the covariant
derivative of fields transforming under the adjoint of $h \left( x \right)$,
$\nabla_{\mu} \cdot = \partial_{\mu} \cdot + \left[ \Gamma_{\mu}, \cdot
\right]$ {\cite{Ecker:1989te}}, with $\Gamma_{\mu}$ as defined in
(\ref{Gammamu}). Assuming now $N_f = 3$, we get
\begin{eqnarray}
  &  & - \frac{1}{4 g_5^2}  \bigintlim_{l_0}^{l_1} \bignone \mathd z \sqrt{g}
  g^{\mu \rho} g^{\nu \sigma}  \left\langle R_{\mu \nu} \left( z \right)
  R_{\rho \sigma} \left( z \right) + L_{\mu \nu} \left( z \right) L_{\rho
  \sigma} \left( z \right) \right\rangle \nonumber\\
  & = & L_1  \left\langle u_{\mu} u^{\mu} \right\rangle^2 + L_2  \left\langle
  u_{\mu} u_{\nu} \right\rangle  \left\langle u^{\mu} u^{\nu} \right\rangle +
  L_3  \left\langle u_{\mu} u^{\mu} u_{\nu} u^{\nu} \right\rangle \nonumber\\
  & - & \mathi L_9  \left\langle f_{+ \mu \nu} u^{\mu} u^{\nu} \right\rangle
  + \frac{1}{4} L_{10}  \left\langle f_{+ \mu \nu} f_+^{\mu \nu} - f_{- \mu
  \nu} f^{\mu \nu}_- \right\rangle \nonumber\\
  & + & \frac{1}{2} H_1  \left\langle f_{+ \mu \nu} f_+^{\mu \nu} + f_{- \mu
  \nu} f^{\mu \nu}_- \right\rangle + \text{terms involving resonances} . 
  \label{p4}
\end{eqnarray}
We have already used the same names as in {\cite{Gasser:1985gg}} for the
coefficients of the operators. However, in this model, they are not unknown,
but rather predicted in terms of the two free parameters of the model. The
expressions for a generic metric are given in (\ref{L1}-\ref{L9+L10}).

\section{Resonance wave-functions} \label{AB}

The EOM for the resonances (\ref{EOM}) becomes, when using the conformally flat
coordinates (\ref{2.2})
\begin{eqnarray}
  - \frac{1}{w \left( z \right)} \partial_z \left( w \left( z \right)
  \partial_z \right) \varphi^X_n & = & M_{X_n}^2 \varphi_n^X,  \label{EOMw}
\end{eqnarray}
where $X = V, A$. Vector and axial cases only differ by the BCs. For the
vector case,
\begin{eqnarray}
  \left. \varphi^V_n \left( z \right) \right|_{z = l_0} & = & 0, 
  \label{phiVUV}\\
  \left. \partial_z \varphi_n^V \left( z \right) \right|_{z = l_1} & = & 0, 
  \label{phiVIR}
\end{eqnarray}
and for the axial one $\left. \varphi_n^A \left( z \right) \right|_{z = l_0} =
\left. \varphi_n^A \left( z \right) \right|_{z = l_1} = 0$.

We normalize the KK wave-functions $\varphi_n$ by imposing on them
\begin{eqnarray}
  \frac{2}{g_5^2}  \bigintlim^{l_1}_{l_0} \mathd z \sqrt{g} g^{\mu \nu}
  g^{\rho \sigma} \varphi^V_m \varphi^V_n & = & \delta_{m n} \eta^{\mu \nu}
  \eta^{\rho \sigma} .  \label{norm}
\end{eqnarray}
With this normalization, one obtains the following quadratic terms for the 4D
KK modes
\begin{eqnarray}
  &  & - \frac{1}{2 g_5^2}  \bigintlim^{l_0}_{l_1} \mathd z \sqrt{g} g^{\mu
  \nu}  \left\{ g^{\rho \sigma}  \left\langle \left( \partial_{\mu} V_{\rho} -
  \partial_{\rho} V_{\mu} \right)  \left( \partial_{\nu} V_{\sigma} -
  \partial_{\sigma} V_{\nu} \right) \right\rangle + 2 g^{55}  \left\langle
  \partial_5 V_{\mu} \partial_5 V_{\nu} \right\rangle \right\} \nonumber\\
  & = & - \frac{1}{4} \eta^{\mu \nu}  \sum_{n=1}^{\infty} \left( \eta^{\rho \sigma} 
  \left\langle \left( \partial_{\mu} V^{\left( n \right)}_{\rho} -
  \partial_{\rho} V^{\left( n \right)}_{\mu} \right)  \left( \partial_{\nu}
  V^{\left( n \right)}_{\sigma} - \partial_{\sigma} V^{\left( n \right)}_{\nu}
  \right) \right\rangle - 2 M_{V_n}^2  \left\langle V_{\mu}^{\left( n \right)}
  V^{\left( n \right)}_{\nu} \right\rangle \right) . 
\end{eqnarray}
The second term in the right-hand side of the above is the sum of the mass
terms for the KK modes.

For the flat case, i.e. $w \left( z \right) = 1$, the solutions to the EOMs
for resonances (\ref{EOMw}) with BCs appropriate to the vector and axial case
 are harmonic functions
\begin{eqnarray}
  \varphi_n^{V, A} \left( z \right) & = & \frac{1}{g_4} \sin \left( M_{V_n,
  A_n} z \right) ,
\end{eqnarray}
where we have used the normalization (\ref{norm}). The masses are given in
Table \ref{TMn}.

For the warp factor $w \left( z \right) = l_0 / z$, corresponding to the RS1
case, the general solution to the EOMs for resonances is written in terms of
Bessel functions
\begin{eqnarray}
  \varphi^X_n \left( z \right) & = & \frac{z}{g_4 N_n^X}  \left( J_1 \left(
  M_{X_n} z \right) + \alpha_n^X Y_1 \left( M_{X_n} z \right) \right), 
  \label{phinV}
\end{eqnarray}
with the dimensionful normalization factor
\begin{eqnarray}
  \left( N_n^X \right)^2 & = & 2 \bigintlim^{l_1}_{l_0} \mathd zz \left( J_1
  \left( M_{X_n} z \right) + \alpha^X_n Y_1 \left( M_{X_n} z \right) \right)^2
  . 
\end{eqnarray}
The relevant BCs for the vector field (\ref{phiVUV}-\ref{phiVIR}) impose
\begin{eqnarray}
  \alpha_n^V & = & - \frac{J_1 \left( M_{V_n} l_0 \right)}{Y_1 \left( M_{V_n}
  l_0 \right)}, \\
  \frac{J_0 \left( M_{V_n} l_1 \right)}{Y_0 \left( M_{V_n} l_1 \right)} & = &
  - \frac{J_1 \left( M_{V_n} l_0 \right)}{Y_1 \left( M_{V_n} l_0 \right)}, 
  \label{=JV}
\end{eqnarray}
while for the axial case, we find
\begin{eqnarray}
  \alpha_n^A & = & - \frac{J_1 \left( M_{A_n} l_0 \right)}{Y_1 \left( M_{V_n}
  l_0 \right)}, \\
  \frac{J_1 \left( M_{A_n} l_1 \right)}{Y_1 \left( M_{A_n} l_1 \right)} & = &
  \frac{J_1 \left( M_{A_n} l_0 \right)}{Y_1 \left( M_{A_n} l_0 \right)} . 
  \label{=JA}
\end{eqnarray}
This allows us to determine numerically both the masses $M_{V_n}, M_{A_n}$ and the
corresponding coefficients  $\alpha_n^{V, A}$ in the linear combinations. For resonances sufficiently light compared to the regulator $1 /
l_0$, or assuming that the regulator is removed, one can obtain approximate
expressions since the coefficient fixing the linear combination is small
\begin{eqnarray}
  \alpha^{V, A}_n & \simeq & \frac{\mathpi}{4} ( M_{V_n, A_n} l_0 )^2
  \hspace{1em} \ll \hspace{1em} 1 . 
\end{eqnarray}
Plugging this into the expression for the wave-functions (\ref{phinV}), one
finds that the masses are approximately determined by the zeroes of the Bessel
functions
\begin{eqnarray}
  J_0 \left( M_{V_n} l_1 \right) & \simeq & 0, \\
  J_1 \left( M_{A_n} l_1 \right) & \simeq & 0 . \nonumber
\end{eqnarray}
This can be solved for numerically, or alternatively, one can obtain the
approximate expressions given in Table \ref{TMn}.

\section{Resonance couplings} \label{AC}

The terms involving resonances in (\ref{p4}) can be written as in
{\cite{Prades:1993ys}} generalized with an index $n$
\begin{eqnarray}
  &  & - \frac{1}{4 g_5^2} \int \mathd z \sqrt{g} g^{\mu \nu} g^{\rho \sigma}
  \left\langle R_{\mu \rho} \left( z \right) R_{\nu \sigma} \left( z \right) +
  L_{\mu \rho} \left( z \right) L_{\nu \sigma} \left( z \right) \right\rangle
  \nonumber\\
  & \supset & - \frac{1}{2 \sqrt{2}}  \sum_{n = 1}^{\infty} \left\{ f_{V_n} 
  \left\langle \left( \nabla_{\mu} V^{\left( n \right)}_{\nu} - \nabla_{\nu}
  V^{\left( n \right)}_{\mu} \right) f_+^{\mu \nu} \right\rangle + \mathi
  g_{V_n}  \left\langle \left( \nabla_{\mu} V^{\left( n \right)}_{\nu} -
  \nabla_{\nu} V^{\left( n \right)}_{\mu} \right)  \left[ u^{\mu}, u^{\nu}
  \right] \right\rangle \right\} \nonumber\\
  & + &  \sum_{n = 1}^{\infty} \left\{ - \frac{1}{2 \sqrt{2}} f_{A_n} 
  \left\langle \left( \nabla_{\mu} A^{\left( n \right)}_{\nu} - \nabla_{\nu}
  A^{\left( n \right)}_{\mu} \right) f_-^{\mu \nu} \right\rangle + \mathi
  \alpha_{A_n}  \left\langle \left[ A^{\left( n \right)}_{\mu}, u_{\nu}
  \right] f_+^{\mu \nu} \right\rangle \right\} + \ldots  \label{5.10}
\end{eqnarray}
where we have only singled out some of the interaction terms. The expressions
for the resonance couplings to sources $f_{V_n}, f_{A_n}$ are
\begin{eqnarray}
  \eta^{\mu \nu} \eta^{\rho \sigma} f_{V_n} & = & \frac{\sqrt{2}}{g_5^2} 
  \bigintlim_{l_0}^{l_1} \mathd z \sqrt{g} g^{\mu \nu} g^{\rho \sigma}
  \varphi_n^V,  \label{fV}\\
  \eta^{\mu \nu} \eta^{\rho \sigma} f_{A_n} & = & \frac{\sqrt{2}}{g_5^2} 
  \bigintlim_{l_0}^{l_1} \mathd z \sqrt{g} g^{\mu \nu} g^{\rho \sigma} \alpha
  \varphi_n^A .  \label{fA}
\end{eqnarray}
The coupling of one vector resonance to two pions is given by
\begin{eqnarray}
  \eta^{\mu \nu} \eta^{\rho \sigma} g_{V_n} & = & \frac{\sqrt{2}}{2 g_5^2} 
  \bigintlim_{l_0}^{l_1} \bignone \mathd z \sqrt{g} g^{\mu \nu} g^{\rho
  \sigma}  \left( 1 - \alpha^2 \right) \varphi_n^V .  \label{gV}
\end{eqnarray}
For the decay constants, i.e. terms that contain a quadratic coupling between
a source and a field, one can obtain expressions involving derivatives at the
UV boundary, rather than bulk integrals. This is done using integration by
parts as well as the EOMs and BCs for both $\alpha$ and $\varphi_n$. The
expressions for the resonance decay constants $f_{V_n, A_n}$ may then be
recast in the form
\begin{eqnarray}
  \eta^{\mu \nu} f_{V_n} M^2_{V_n} & = & \left. - \frac{\sqrt{2}}{g_5^2} 
  \sqrt{g} g^{\mu \nu} g^{55} \partial_5 \varphi^V_n \right|_{l_0}, 
  \label{FV} \\
  \eta^{\mu \nu} f_{A_n} M^2_{A_n} & = & \left. - \frac{\sqrt{2}}{g_5^2} 
  \sqrt{g} g^{\mu \nu} g^{55} \partial_5 \varphi^A_n \right|_{l_0} .  
  \label{FA}
\end{eqnarray}
These expressions (\ref{FV}-\ref{FA}) have a more holographic twist than
(\ref{fV}-\ref{fA}), since they give the couplings of KK modes as boundary
values of derivatives with respect to the fifth coordinate of the
wave-functions, rather than in terms of bulk integrals.

Note as well that a similar equivalence can be proved for the pion decay
constant, using again integration by parts to rewrite (\ref{fpi}) as
\begin{eqnarray}
  \eta^{\mu \nu} f^2 & = & \left. \frac{1}{g_5^2}  \sqrt{g} g^{\mu \nu} g^{55}
  \partial_5 \alpha \right|_{l_0} .  \label{fpi2}
\end{eqnarray}
\section{5D regularization of infinite sums} \label{AD}

The 5D origin of the model means that the different resonance couplings are
given by overlap integrals of wave-functions $\varphi_n \left( z \right)$,
which obey completeness relations {\footnote{This relation is valid when the
distribution is applied to a function satisfying the same BCs as
$\varphi^V_{_n}$. For functions that do {\tmem{not}} satisfy these BCs, one
obtains additional boundary terms, which would be missed by blindly applying
(\ref{kpl}).}}
\begin{eqnarray}
  \frac{2}{g_5^2}  \sqrt{g} g^{\mu \nu} g^{\rho \sigma}  \sum_{n=1}^{\infty} \varphi_n^V
  \left( z \right) \varphi_n^V \left( z' \right) & = & \eta^{\mu \nu}
  \eta^{\rho \sigma} \delta \left( z - z' \right) .  \label{kpl}
\end{eqnarray}
Using this remark, we were able to derive the sum rules of Section \ref{SR},
which stipulate that sums of products of resonance couplings are related to GB
interactions. The sums (\ref{fVgV}-\ref{KSFRsum}) were convergent, so no problems
were encountered there. We now focus on the sums $\sum_n f_{V_n}^2$ and
$\sum_n f_{A_n}^2$, which are logarithmically divergent for the case of the
RS1 metric in the limit $l_0 \longrightarrow 0$, see Table \ref{Tn}.

In principle, one has to relate the two regulators for the infinite sums in a
way that preserves chiral symmetry, as advocated in {\cite{Golterman:2002mi}}.
The usual 4D procedure involves cutting the infinite sums over $n$, but chopping
the KK tower would be inconsistent with 5D gauge invariance. However, we can
rewrite the sums as integrals over the fifth coordinate, obtaining
\begin{eqnarray}
  \sum_{n = 1}^{\infty} f_{A_n}^2 & = & \frac{1}{g_5^2} 
  \bigintlim_{l_0}^{l_1} \mathd zw \left( z \right) \alpha^2 \left( z \right),
  \label{fA2}
\end{eqnarray}
for the axial case and (\ref{logdiv}) for the vector one. The
extra-dimensional setup yields a clear prescription which preserves the
$\tmop{SU} \left( N_f \right) \times \tmop{SU} \left( N_f \right)$ symmetry.
It involves combining $\sum_n f_{V_n}^2 \bignone$ (\ref{logdiv}) and $\sum_n
f_{A_n}^2 \bignone$ (\ref{fA2}) as
\begin{eqnarray}
  \sum_{n = 1}^{\infty} f_{V_n}^2 - \sum_{n = 1}^{\infty} f_{A_n}^2 & = &
  \frac{1}{g_5^2}  \bigintlim_{l_0}^{l_1} \mathd z w \left( z \right)  \left( 1
  - \alpha^2 \left( z \right) \right),  \label{int(1-alpha2)}
\end{eqnarray}
and sending the regulator $1 / l_0$ to infinity {\tmem{afterwards}}. Since the
BC (\ref{alphaUV}) imposes $\alpha \left( l_0 \right) = 1$, the result is
finite
\begin{eqnarray}
  \sum_{n = 1}^{\infty} \left( f_{V_n}^2 - f_{A_n}^2 \right) & = & - 4 L_{10}
  .  \label{fV2-fA2}
\end{eqnarray}
This expression generalizes the one known for the case of a finite number of
resonance. The result (\ref{fV2-fA2}) is also valid for any other metric where
the integral in the right-hand side of (\ref{int(1-alpha2)}) converges: for
flat space in particular, each one of the two sums is separately convergent.

For RS1, the method described above is the appropriate way to handle the
divergence, since vector and axial fields live in the {\tmem{same}} extra
dimension, and both come from linear combinations of the original $L_{\mu}$
and $R_{\mu}$ fields. Rewriting the sums over $n$ as integrals over the bulk
with the regulator $1 / l_0$ present takes this into account, while preserving
the $\tmop{SU} \left( N_f \right) \times \tmop{SU} \left( N_f \right)$
symmetry. Removing the regulator then yields the result (\ref{fV2-fA2}).

The prescription is clearly dictated by the fact that the model is defined in
an extra-dimensional setup. As a matter of fact, working in 5D, none of these subtleties
would be necessary. The result is transparent and no ordering in taking the
limits is necessary since the only scales are the momentum $q$ and the
position of the IR brane $l_1$. No intermediate KK description is needed: in fact, the  description in terms of resonances hides the local structure of the fifth dimension. In order to preserve this
structure, one can still use the 4D representation in terms of sums,
albeit with a regularization appropriate to 5D. One example is the
so-called {\tmem{KK regularization}} of {\cite{Delgado:2001ex}}, which we consider in Section \ref{KKreg}.

We are going to see how one obtains the two Weinberg sum rules (WSRs) and
their generalizations by considering moments of the $\Pi_{L R}$ correlator
\begin{eqnarray}
  \text{for } k \geqslant 1, \hspace{2em} \frac{1}{\mathpi} 
  \bigintlim^{\infty}_0 \mathd ss^{k - 1} \tmop{Im} \Pi_{L R} \left( s \right)
  & = & \sum_{n = 1}^{\infty} f_{V_n}^2 M_{V_n}^{2 k} - \sum_{n = 1}^{\infty}
  f_{A_n}^2 M_{A_n}^{2 k} - f^2 \delta^{k 1} \bignone .  \label{WSRsn}
\end{eqnarray}
Note that the case of $L_{10}$ studied above corresponds to
$k = 0$ with the pion pole removed. The WSRs are closely related to the values
of the OPE coefficients in $\Pi_{L R}$ (\ref{noLOP}): we will find that the
sums (\ref{WSRsn}) vanish, as expected from (\ref{noLOP}).

\subsection{KK regularization}
\label{KKreg}

Limiting ourselves to flat space, we assume that the (electroweak) matter
content is spread with a Gaussian distribution, rather than being localized as
a delta function on the UV brane.  The spread of the Gaussian in the fifth
coordinate would be proportional to a length $\sqrt{\tau}$. The (electroweak)
decay constants are then expected to be modified as
\begin{eqnarray}
  f_{V_n}^2 & \rightarrow & f_{V_n}^2 \mathe^{- \tau M_{V_n}^2},  \label{tau}
\end{eqnarray}
and similarly for the axial case. The regulator $\tau$ should be sent to zero
at the end of the computations.

We now directly define the separate regulated sums representing the vector and
axial contributions to the $k$-th moment of (\ref{WSRsn})
\begin{eqnarray}
  V^{\left( k \right)} \left( \tau \right) & \equiv & \sum_{n = 1}^{\infty}
  f_{V_n}^2 \mathe^{- \tau M_{V_n}^2} M_{V_n}^{2 k},  \label{Vktau}\\
  A^{( k )} \left( \tau \right) & \equiv & \sum_{n = 1}^{\infty} f_{A_n}^2
  \mathe^{- \tau M_{A_n}^2} M_{A_n}^{2 k} .  \label{Aktau}
\end{eqnarray}
This regulation makes both sums finite for {\tmem{any}} $k$: the powers of
masses are compensated by the exponential suppression for any $\tau > 0$. We next introduce the
variable
\begin{eqnarray}
  q & \equiv & \mathe^{- \tau \left( \frac{\pi}{l_1} \right)^2}, 
\end{eqnarray}
which tends to one as the regulator $\tau$ is sent to zero. We rewrite the sums (\ref{Vktau}-\ref{Aktau}) in terms of elliptic
theta functions as
\begin{eqnarray}
  V^{\left( k \right)} \left( \tau \right) & = & \frac{1}{\left( g_4 l_1
  \right)^2}  \left( \frac{\mathpi}{l_1} \right)^{2 \left(k-1\right)}  \left[ q
  \frac{\partial}{\partial q} \right]^{k-1} \vartheta_2 \left( 0, q \right), \\
  A^{( k )} \left( \tau \right) & = & \frac{1}{\left( g_4 l_1 \right)^2} 
  \left( \frac{\mathpi}{l_1} \right)^{2 \left(k-1\right)}  \left[ q \frac{\partial}{\partial
  q} \right]^{k-1} \left\{ \vartheta_3 \left( 0, q \right) - 1 \right\} . 
\end{eqnarray}
The separate sums have been regulated, and only diverge if $q$ is set to 1. We
can however first combine the well-defined vector and axial sum in order to
examine the moments of $\Pi_{L R}$ introduced in (\ref{WSRsn}). We use
\begin{eqnarray}
  \vartheta_3 \left( 0, q \right) - \vartheta_2 \left( 0, q \right) & = &
  \vartheta_4 \left( 0, q^{1 / 4} \right), 
\end{eqnarray}
and the fact that  $\vartheta_4 \left( 0, q \right)$ and all its derivatives vanish
as $q \longrightarrow 1$
\begin{eqnarray}
  \text{for } k \geqslant 0 \hspace{2em} \frac{\partial^k}{\partial q^k} 
  \vartheta_4 \left( 0, q \right)  & \underset{q
  \rightarrow 1}{\longrightarrow} & 0 . 
\end{eqnarray}
With all these rewritings, we obtain that the finite expressions combine to
give the expected results in the limit where the regulator is removed
\begin{eqnarray}
  V^{\left( 1 \right)} \left( \tau \right) - A^{\left( 1 \right)} \left( \tau
  \right) & \underset{\tau \rightarrow 0}{\longrightarrow} & f^2, \\
  \text{for} k \geqslant 2, \hspace{2em} V^{\left( k \right)} \left( \tau
  \right) - A^{\left( k \right)} \left( \tau \right) & \underset{\tau
  \rightarrow 0}{\longrightarrow} & 0 . 
\end{eqnarray}
In other words, we reobtain the infinite number of generalized Weinberg sum
rules. The 5D regularization has automatically modified {\tmem{both}} the
vector and axial contributions to yield this result: regulating in
$x^5$-position space affects both $L_{\mu}$ and $R_{\mu}$ gauge fields, and
hence $V_{\mu}$ and $A_{\mu}$ in a definite way, so as to preserve the
$\tmop{SU} \left( N_f \right) \times \tmop{SU} \left( N_f \right)$ symmetry.

\subsection{Generalized Weinberg sum rules}

Having seen that the  axial and vector sums separately diverged, but
that the difference could yield a meaningful result, we now work with
a generic metric. If we do not insist on regulating the delta
functions, we can use the completeness relation (\ref{kpl}) to rewrite
the sums in (\ref{WSRsn}) as expressions in $x^5$
position-space. These expressions involve divergent boundary terms on the UV brane. The divergences will again cancel between vector and axial contributions.
Using the expressions for the
resonance couplings and the completeness relations, we find the
following result corresponding to the first WSR
\begin{eqnarray}
  \sum_{n = 1}^{\infty} f_{V_n}^2 M_{V_n}^2 - \sum_{n = 1}^{\infty} f_{A_n}^2
  M_{A_n}^2 & = & f^2 . 
\end{eqnarray}
For moments with $k \geqslant 2$ in (\ref{WSRsn}), we find that the model satisfies an infinite number of generalized WSRs, i.e.
\begin{eqnarray}
  \text{for } k \geqslant 2, \hspace{2em} \sum_{n = 1}^{\infty} f_{V_n}^2
  M_{V_n}^{2 k} - \sum_{n = 1}^{\infty} f_{A_n}^2 M_{A_n}^{2 k} & = & 0 . 
  \label{GWSRs}
\end{eqnarray}
Taken separately, the 5D expressions for each sum would involve terms localized in the
extra dimension: delta functions signaling that our world is living in $l_0$.
Those divergent contributions cancel between the vector and axial currents.
The fact that vector and axial resonances descend from 5D fields living in the
{\tmem{same}} extra-dimension means that the result respects chiral symmetry.

A last comment is in order: a delta function distribution in any 5D theory
must be taken as a regularized delta function. This would give a finite value
at intermediate stages for both axial and vector sums, as exemplified in the
KK-regularization used in Section \ref{KKreg}.

\end{appendix}

\nocite{B,Gif:1996,Maiani:1995ve}
\bibliography{bib-biblio.bib}

\bibliographystyle{JHEP}

\end{document}